\documentclass[aps,prb,floatfix,amsmath,amssymb,preprint,eqsecnum,nofootinbib,superscriptaddress]{revtex4}
\usepackage{graphicx}
\usepackage{dcolumn}
\usepackage{bm}
\usepackage{amsmath, amsfonts, amssymb}
\usepackage{amsxtra}
\usepackage{amsthm}

\usepackage{soul} 

\newcommand{\bdag}{\hat{b}^\dag}

\renewcommand{\ddag}{\hat{d}^\dag}

\newcommand{\bra}[1]{\left\langle#1\right|}
\newcommand{\ket}[1]{\left|#1\right\rangle}

\newcommand{\abs}[1]{\left|#1\right|}
\newcommand{\ave}[1]{\langle#1\rangle}

\newcommand{\eqnref}[1]{Eq.~(\ref{#1})} 
\newcommand{\fref}[1]{Fig.~\ref{#1}}    
\newcommand{\secref}[1]{Section~\ref{#1}}

\newcommand{\beq}{\begin{equation}}
\newcommand{\eeq}{\end{equation}}
\newcommand{\ba}{\begin{array}{ccc}}
\newcommand{\ea}{\end{array}}

\def\bea{\begin{eqnarray}}
\def\eea{\end{eqnarray}}

\usepackage{setspace} 
\begin{document}

\title{Frustrated quantum Ising spins simulated\\ by spinless bosons in a tilted lattice:\\
from a quantum liquid to antiferromagnetic order}

\author{Susanne Pielawa}
\affiliation{Department of Condensed Matter Physics, Weizmann Institute of Science, Rehovot, 76100, Israel}
\author{Erez Berg}
\affiliation{Department of Condensed Matter Physics, Weizmann Institute of Science, Rehovot, 76100, Israel}
\affiliation{Department of Physics, Harvard University, Cambridge MA
02138}
\author{Subir Sachdev}
\affiliation{Department of Physics, Harvard University, Cambridge MA
02138}

\date{\today \\
\vspace{1.6in}}

\begin{abstract}
We study spinless bosons in a decorated square lattice with a
near-diagonal tilt. The resonant subspace of the
tilted
Mott insulator is
described by an effective Hamiltonian of frustrated quantum Ising
spins on a non-bipartite lattice. This generalizes an earlier
proposal for the unfrustrated quantum Ising model in one dimension
which was realized in a recent experiment on ultracold $^{87}$Rb
atoms in an optical lattice. Very close to 
diagonal tilt,
we find a quantum liquid state which is continuously connected to
the paramagnet. Frustration can be reduced by increasing the tilt
angle away from the diagonal, and the system undergoes a
transition to an antiferromagnetically ordered state. Using
quantum Monte Carlo simulations and exact diagonalization, we find
that for realistic system sizes the antiferromagnetic order 
appears to be quasi-one-dimensional; however, in the thermodynamic
limit the order is two-dimensional.
\end{abstract}
\maketitle

\section{Introduction}

Recent experimental progress in the field of ultracold atomic gases
has made it possible to study quantum many-body physics in a controllable
and clean setting. This makes cold atoms in optical lattices candidates
for analog quantum simulators of real materials \cite{PhysRevLett.81.3108,RevModPhys.80.885, Bloch2012,Cirac2012}. Since the observation
of a quantum phase transition from a superfluid state to an interaction
driven insulating state\cite{Greiner2002}, there has been much
effort to simulate other correlated quantum phases, such as magnetic
phases. Many proposals suggest using an internal degree of freedom
of the atoms to simulate a spin degree of freedom\cite{1367-2630-5-1-113, PhysRevLett.90.100401, PhysRevA.84.031603}. 
Virtual hopping
processes then lead to an effective magnetic interaction called superexchange \cite{AssaBook, Trotzky18012008}.
The energy scale of those processes is still low compared to experimentally
reachable temperatures, and so magnetic long range order has not yet
been observed.

An important milestone was recently reached taking a surprising new
route: an equilibrium quantum phase transition of an antiferromagnetic
spin chain was simulated using spinless bosons in a non-equilibrium
situation. Following the theoretical proposal of
Refs.~\onlinecite{SachdevMI} and \onlinecite{PhysRevB.69.075106}, Simon {\em et al.}\cite{GreinerIsing} examined
a one-dimensional array of $^{87}$Rb atoms
in an optical lattice; an additional potential gradient (\textquoteleft{}tilt\textquoteright{})
drove the transition from the Mott insulating state to a state with density wave order. This happened
in a metastable state, which is not the ground state of the full bosonic
hamiltonian. However, the dynamics of the tilted lattice was confined
to a resonantly connected effective subspace, which has an energy
bounded from below, and so a mapping
to an antiferromagnetic Ising model in a transverse and longitudinal
field is possible.\cite{SachdevMI}
Changing the tilt magnitude corresponds to changing the strength of the longitudinal field, and this takes the systems through an Ising quantum phase transition, 
which was observed using single site resolutions\cite{bakrgreiner2, Sherson2010, Gemelke2009,Gericke2008,Nelson2007}. The dynamics of the one-dimensional system has also been studied theoretically
\cite{PhysRevA.69.053616, 1ddyn1, 1ddyn2}. 

\begin{figure}[tb]
\begin{center}
\includegraphics[width=.75\textwidth]{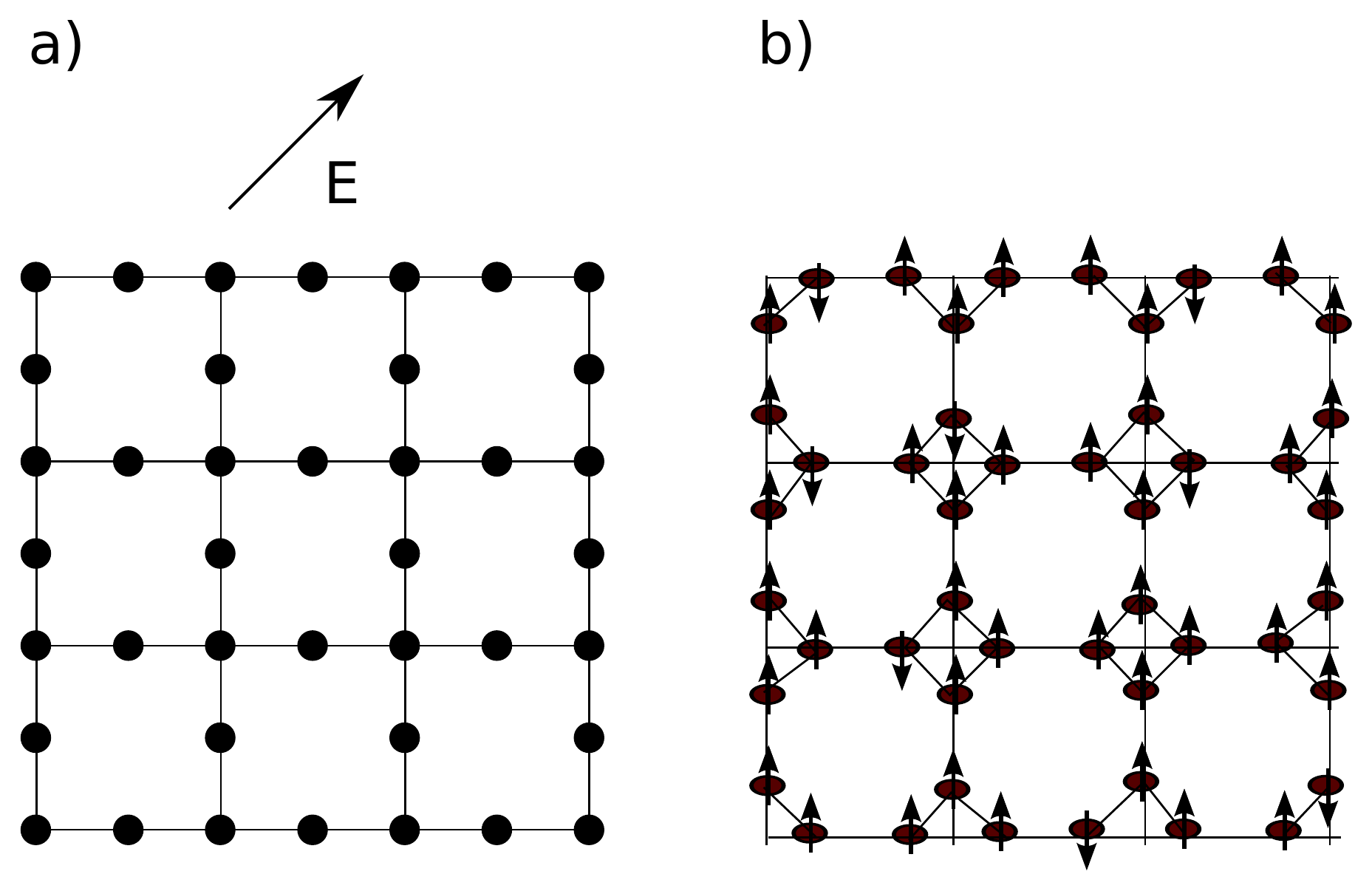}
\end{center}
\caption{Decorated square lattice in a near-diagonal tilt: a) shows the physical lattice for the bosons. The spins reside on the links, which form the lattice shown in b).
The effective resonant subspace of the boson model of the decorated square lattice (a) maps to an antiferromagnet on the octagon-square-cross lattice (b) in a strong longitudinal and in a transverse magnetic field. If the tilt is not exactly diagonal, then spins on horizontal lines experience a different longitudinal field than spins on vertical lines. The spin lattice is not bipartite, and in this sense the antiferromagnet is frustrated.
}
\label{spinlattice}
\end{figure}

In Ref.~\onlinecite{PielawaTiltMI}, we have shown that a variety of correlated
phases are possible in tilted two dimensional lattices.
A mapping to a spin model is in general not possible in two dimensions.
In this paper we focus on a lattice- and tilt configuration where a spin mapping is possible also in two dimension: a diagonally tilted decorated square lattice of bosons leads to a spin model on a non-bipartite lattice, the ``octagon-square-cross lattice'' (see Figure \ref{spinlattice}).
We start from the Mott insulator with a filling factor of one atom
per lattice site, and assume that effective three-body interaction
are important, 
such that triply occupied sites are not allowed. It has been
shown\cite{PielawaTiltMI} that in this system, three-body
interactions have important qualitative effects; in particular, if
three-body interaction are negligible, then the system maps to a
quantum dimer model on a square lattice, as discussed in Ref. \onlinecite{PielawaTiltMI}.

When the potential drop per lattice site is comparable to the on-site repulsion, then the only processes allowed in the resonant subspace are creations of `dipoles' along the links of the lattice. A dipole is created when a boson follows the tilt direction and moves onto a neighboring site, which already contains one boson. This process costs the on-site repulsion energy $U$ and gains potential energy $E$. In the parameter regime $\abs{U-E}\ll U,E$ all other processes are off-resonant.\cite{SachdevMI}

Due to conservation of energy any lattice site can be part of no
more than one dipole.
%
We can map the Hamiltonian of the resonant subspace to a spin model by associating a spin state to each link:
spin up if no dipole has been created on that link,
and spin down if a dipole has been created on that link.
The hard constraint forbidding overlapping dipoles translates to a strong antiferromagnetic interaction in a strong longitudinal field, and so we obtain an Ising antiferromagnet.

As this lattice is not bipartite, antiferromagnetic order is not possible, even for a weak longitudinal field (where the antiferromagnetic interaction dominates over the magnetic field).  In this sense the Ising spin model on this lattice is geometrically frustrated.

\begin{figure}[tb]
\begin{center}
\includegraphics[width=\textwidth]{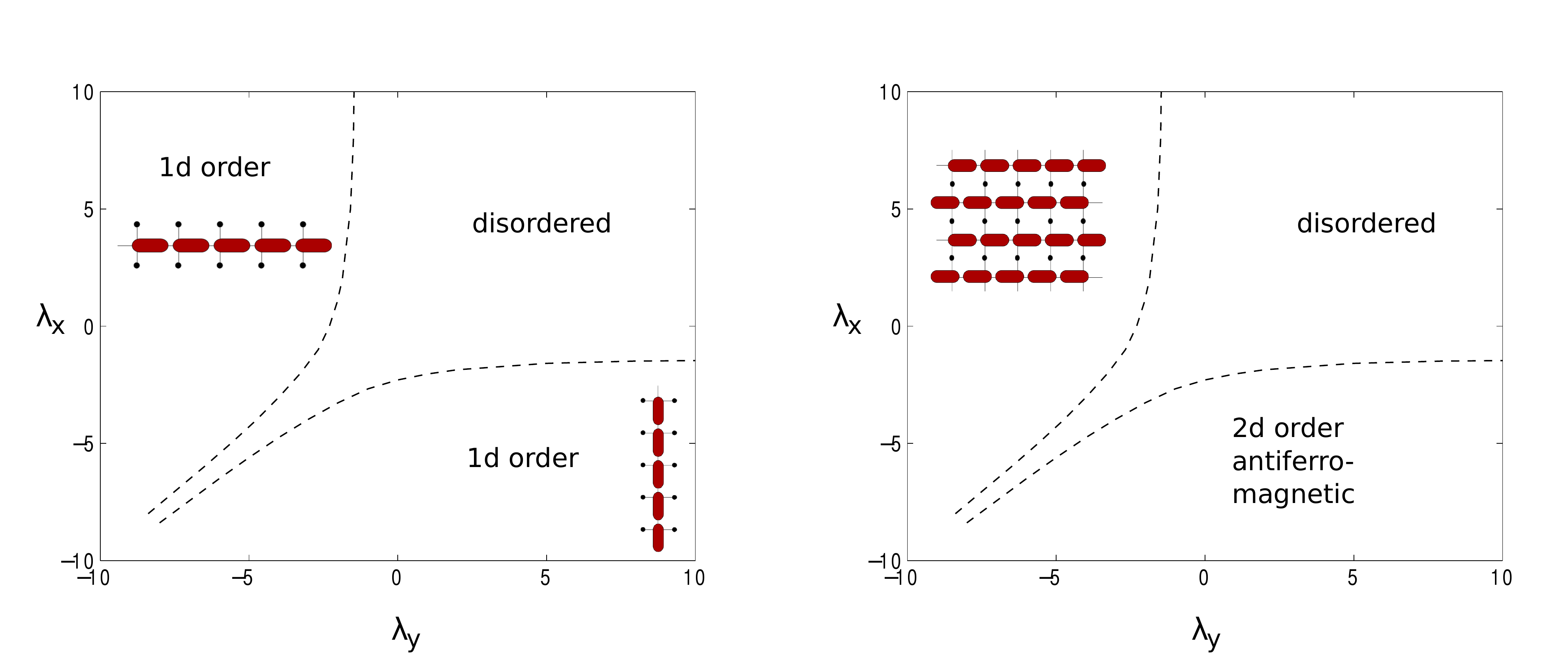}
\end{center}
\caption{Phase diagram of the near-diagonally tilted decorated
square lattice, determined by quantum Monte-Carlo calculations.
The tuning parameters are
$\lambda_x$ and $\lambda_y$ which parametrize the cost for having
a dipole in $x$ and $y$ direction, respectively. a) for realistic
system sizes there appears to be only one-dimensional order.
This is a finite-size artifact, 
arising from the fact that for these system sizes the finite-size
gap dominates over the (antiferromagnetic) inter-chain coupling.
 b) in the thermodynamic limit we predict 2d quantum Ising transitions to a phase where neighboring chains are aligned antiferromagnetically.
}
\label{schematic_phase_diag}
\end{figure}

Let us first briefly review\cite{PielawaTiltMI} the case of a diagonal tilt.
The parent Mott insulator remains stable to a weak tilt.
In the other limit, the strongly tilted case, the system wants to maximize the number of dipoles.
Due to the nearest-neighbor exclusion constraint, and due to the lattice geometry, there can be no more than one dipole per unit cell. The lattice has four links per unit cell, and so there is a lot of room for the dipoles to fluctuate. Quantum fluctuations then create a unique and gapped ground state: an equal amplitude superposition of all classically allowed dipole coverings~\cite{PielawaTiltMI}. This disordered quantum liquid state is continuously connected to the parent Mott insulator; it is part of the same phase, as shown in the phase
diagram in Fig.~\ref{schematic_phase_diag}.

The physics becomes more interesting when the tilt $\vec E = (E_x, E_y)$ deviates slightly from the diagonal, $E_x\ne E_y$.
Dipoles in $x$ direction and dipoles in $y$ direction now do not cost the same energy, and
we define $\Delta_{x}=U-E_{x}$ ($\Delta_y=U-E_y$) the energy associated with a dipole in $x$ ($y$) direction.

In the limit $\Delta_y\rightarrow \infty$; $\Delta_x / \Delta_y \rightarrow 0$ the system can reduce its potential energy by maximizing the number of dipoles in $y$ direction; no dipoles in $x$ direction are created in the ground state.
Thus the system decouples into a collection of horizontally aligned chains.
In the absence of vertical dipoles these chains cannot interact with each other, and so each of them undergoes an independent 1d Ising transition as a function of $\Delta_x$ (see Fig.~\ref{schematic_phase_diag}).

In this paper we study the full phase diagram of the near-diagonally tilted decorated square lattice. An important question is whether there  is a region in the phase diagram where 2d order develops, i.e. whether a coupling develops between the above mentioned chains.
We find that for realistic system sizes the crossover looks like the one of a collection of independent 1d chains.
This is due to the fact that the inter-chain coupling is small compared to the finites-size gap of each chain between the symmetric and anti-symmetric superposition of its two ground states. There is thus no 2d order.
A schematic phase diagram for this situation is shown in Figure \ref{schematic_phase_diag}a.
The situation is different in the thermodynamic limit.
As the finite-size gap vanishes the inter-chain coupling dominates, causing the chains to align anitferromagnetically. The schematic phase diagram in the thermodynamic limit is shown in Figure \ref{schematic_phase_diag}a.

The remainder of this paper is organized as follows. In
\secref{sec:Model} we introduce the model and describe the
effective resonant subspace. In \secref{sec:MC} we study the
system by quantum Monte-Carlo (QMC) simulations and find no sign
of a coupling between the chains. In
\secref{sec:inter-chain-coupling} we will show by exact
diagonalization of a model system consisting of only two chains
that there is indeed a very small coupling between these chains,
which arises from processes in very high order in perturbation
theory. We present conclusions in \secref{sec:Conclusions}.

\section{Model}
\label{sec:Model}
In this section we describe the effective resonant subspace of a
near-diagonally tilted decorated square lattice. We begin by
recalling the Hamiltonian of a tilted Mott insulator. It is
described by the generalized bosonic Hubbard model with an
additional potential gradient along a certain direction, $H=H_{\rm
kin}+H_{\rm U}+H_{\rm tilt}$:
\begin{subequations}
\begin{eqnarray}
H_{\rm kin} &=& - t \sum_{<ij>} \left( \bdag_i \hat b_j + \bdag_j \hat b_i \right) \\
H_{\rm U}&=&\frac U 2 \sum_i \hat n_i (\hat n_i-1) + \frac {U_3} {6} \sum_i n_i (n_i-1)(n_i-2)
+\dots \\
\label{eq:repulsion}
H_{\rm tilt}&=&- E \sum_i {\bf e} \cdot {\bf r}_i \, \hat n_i .
\label{hubbard}
\end{eqnarray}
\end{subequations}
Here $\hat b_i$ are canonical boson operators on lattice sites $i$ at spatial co-ordinate
${\bf r}_i$, and $\hat n_i \equiv \bdag_i \hat b_i$.
The first term in $H_{\rm U}$ describes two-body interaction. The
second term is an effective three-body interaction, generated by
virtual processes involving higher bands \cite{multibody,
threebody, three-body-Johnson}. Such a term is present in ultracold atomic systems,
and dramatically changes the physics of a tilted lattice, as we
have shown in Ref. \onlinecite{PielawaTiltMI}: if $U_3$ is not
negligible compared to other energy scales in the problem, then
this term causes processes which create triply occupied sites to
be off-resonant. This effect is independent of the sign of $U_3$. Indeed, in recent cold atom 
experiments $U_3$ has been measured to be negative\cite{multibody, threebody, GreinerPhotoAssistedTunneling}.
The potential gradient is $E$, and the fixed vector
${\bf e}$ is normalized so that the smallest change in potential energy between neighboring lattice sites
has magnitude $E$.
We assume that the potential drop per lattice site $E$ is comparable to the on-site repulsion $U$. The tilt has now two components $\vec E=(E_x, E_y)$, we define $\Delta_x=U-E_x$ and $\Delta_y=U-E_y$, and we work in the parameter regime where
\begin{equation}
\abs{\Delta_x}, \abs{\Delta_y}, t \ll \abs{U}, \abs{E} , \abs{U_3}.
\end{equation}
We assume that the parent Mott insulator has filling factor one atom per lattice site.
The effective resonant subspace of the near-diagonally tilted decorated square lattice is then described by the Hamiltonian
\bea
\hat H&=& \Delta_x \sum_{i\in \textrm{x-links}} \ddag_i \hat d_i + \Delta_y \sum_{j\in \textrm{y-links}} \ddag_j\hat d_j-\sqrt{2}t\sum_{a}\left(\hat d_a +\ddag_a\right),
\label{eq:dipole-model}
\eea
here $\ddag_a$ ($\hat d_a$) creates (annihilates) a dipole on a link $a$;
where the first sum runs only over links aligned in $x$ direction, the second sum only over links aligned in $y$ direction. These two terms describe the energy cost/gain for having a dipole. The last sum comes from the hopping term and describes creation and annihilation of dipoles; it runs over all links. The dipoles obey a hard-core constraint: there can be no more than one dipole on each link. Additionally there is a constraint which does not allow dipoles to overlap on a site: each lattice site can be part of no more than one dipole.

For 
following discussion, 
we define the two independent tuning parameters
\begin{eqnarray}
\lambda_{x}&=&\frac{\Delta_{x}}{\sqrt{2}t}, \qquad
\lambda_{y}=\frac{\Delta_{y}}{\sqrt{2}t}
\end{eqnarray}
These two parameters can take all real values.

\subsection{Description by a constrained five-state model}
\begin{figure}[tb]
\begin{minipage}{.59\textwidth}
\begin{center}
\includegraphics[width=.5\textwidth]{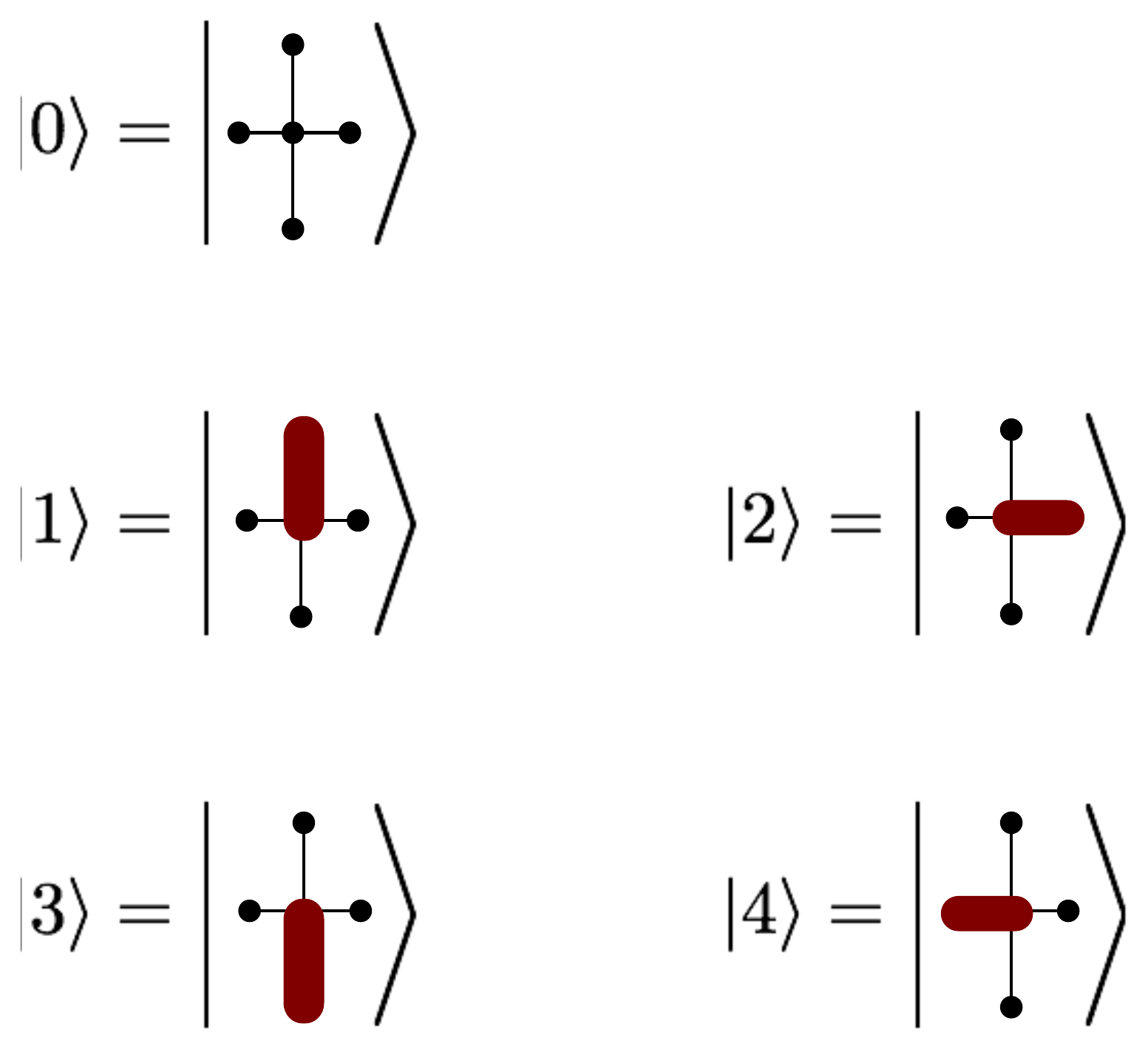}
\end{center}
\end{minipage}
\caption{
Effective five-state model: each unit cell of the decorated square lattice can be in one out of five possible states: having no dipole, or having a dipole on one of the four links. There is also a constraint that dipoles may not overlap: two neighboring unit cells may not have dipoles directed towards each other. Thus the effective model is a constraint five state model on a simple square lattice.
\label{fig:5-state-model}}
\end{figure}
We can describe the resonantly connected subspace for all values of our tuning parameters $\lambda_x$ and $\lambda_y$ by a constrained five-state model on a simple square lattice. We let our unit cell be centered about the sites with four neighbors. Each of the unit cells may be in one out of five states, see \fref{fig:5-state-model}: It may contain no dipole (state $\ket 0$), or it may contain one dipole, and there are four links to choose from (states $\ket 1$, $\ket 2$, $\ket 3$, $\ket 4$). The Hamiltonian of a single site
is then given by
\begin{eqnarray}
H_{{\rm site}} & = & H_{{\rm pot}}+H_{{\rm kin}}\nonumber \\
H_{{\rm pot}} & = & \lambda_{x}\left(\ket 2\bra 2+\ket 4\bra 4\right)+\lambda_{y}\left(\ket 1\bra 1+\ket 3\bra 3\right)\nonumber \\
H_{{\rm kin}} & = & -\ket 0\left(\bra 1+\bra 2+\bra 3+\bra 4\right)+{\rm h.c.}\label{eq:decSq-single-site}.
\end{eqnarray}
Summing over all sites we obtain the free Hamiltonian of the five state system
\[
H_{{\textrm{free}}}=\sum_{x,y=1}^{L_{x},L_{y}}H_{{\rm site}}(x,y).
\]
Additionally there is a constraint that the dipoles may not overlap: two neighboring unit cells may not point toward each other. We take this into account by projecting out all the states which would create such a collision
\[
H^{c}=P_{c}H_{{\rm{free}}}P_{c}
\]
where $P_{c}$ is a projection operator which projects out all the
states that are forbidden by the constraint.

\subsection{Mapping to a frustrated Ising spin model}
\label{sub:spinmapping}
In Refs. ~\onlinecite{SachdevMI} and~\onlinecite{GreinerIsing}, the physics of a tilted one-dimensional Mott Insulator was described by an antiferromagnetic Ising spin chain in a transverse and longitudinal field. In the same spirit, we map the diagonally tilted decorated square lattice to an antiferromagnetic spin model, also in longitudinal and transverse field, on a \emph{frustrated} lattice. Note that the spin degrees of freedom reside on the links of the decorated square lattice, so that the lattice of the spin model is an `octagon-square-cross' lattice as depicted in \fref{spinlattice}.
This lattice has four sites per unit cell, and each spin has $z_{\rm{coord}}=4$ neighbors. As the lattice is not bipartite, an antiferromagnetic spin model on this lattice is frustrated, and two-dimensional Ising 
order is not possible.
The Hamiltonian of the resonant subspace can be described by the following spin model
\begin{subequations}
\begin{eqnarray}
H&=& J\left(\sum_{\ave{i,j}} S_z^i S_z^j -h^{\rm {LR}}_z\sum_{i\in\rm LR} S_z^i
-h^{\rm {UD}}_z\sum_{i\in\rm UD} S_z^i
- h_x \sum_i S^i_x\right)\\
h^{\rm{LR}}_z &=& \left(2-\frac {\Delta_x} J\right), \qquad
h^{\rm{UD}}_z = \left(2-\frac {\Delta_y} J\right)\\
h_x &=& 2 \sqrt{2}\frac { t} J
\end{eqnarray}
\end{subequations}
where $\vec S=\frac 1 2 \vec \sigma$. The second sum ($i\in
\rm{LR}$) is over all spins which reside on lines in horizontal
direction, see \fref{spinlattice}, and the third sum
($i\in\rm{UD}$) is over spins that reside on vertical lines. While
the first three terms all commute with each other, the last term
does not. It is this transverse field which makes this a quantum
problem. The strong antiferromagnetic interaction and the strong
longitudinal field are introduced to realize the constraint:
having two neighboring spin down costs an energy of order $J$. The
mapping becomes exact
in the limit $J\rightarrow\infty$\footnote{
When taking the limit $J\rightarrow\infty$, the parameter $\Delta$ has to be kept fixed, not
$h_\alpha$. The terms proportional to J are then used to realize the
constraint, as explained in Ref. \onlinecite{SachdevMI}
}. As in one
dimension, this is of course not a mapping of the full bosonic
model to a spin model, but of the resonantly connected subspace.

We will phrase most of the following discussion in the language of the constrained five-state model, keeping in mind that the results can directly be applied to the frustrated Ising spin model.

\subsection{Limiting cases}
We understand the system in the following limiting cases
\begin{enumerate}
\item $\lambda_x, \lambda_y\rightarrow \infty$, $\lambda_x/\lambda_y=1$ (weak diagonal tilt):
The parent Mott insulator is stable to a weak tilt, and so the dipole vacuum is the ground state in this limit.
Dipole creation costs a large amount of energy, and so dipoles are
only virtually created. \item $\lambda_x, \lambda_y\rightarrow
-\infty$, $\lambda_x/\lambda_y=1$ (strong diagonal tilt): This is
the quantum liquid state described in Ref.
\onlinecite{PielawaTiltMI}. The number of dipoles is maximized,
and the ground state is an equal amplitude superposition of all
dipole product states that fulfill the constraint. As we have
shown in Ref. \onlinecite{PielawaTiltMI} this is a disordered
state; the ground state is unique and gapped.
\item $\lambda_y\rightarrow +\infty$; $\lambda_x/\lambda_y \rightarrow 0$: along this line vertical dipole states on links aligned in $y$ direction cannot be occupied, as they cost an infinite amount of energy, while horizontal dipoles along links in $x$ direction are accessible. In this limit the system decouples into a collection of horizontal one-dimensional chains. These chains are effectively one-dimensional, they undergo a phase transition in the Ising universality class\footnote{The symmetry which is broken in the ordered phase is a
reflection symmetry. The lattice has a larger translation
symmetry (unless $\Delta_y$ is strictly infinite), which is not broken in the ordered phase.}
at a
critical value of $\lambda_x=-1.31$ (which is the same as in the
one-dimensional case).
\end{enumerate}
We expect the one-dimensional order within each chain to persist when $\lambda_y$ takes on finite values. Neighboring chains may then interact via dipole states aligned in $y$ direction, which might lead to a coupling between these chains, and thus to two-dimensional order. \\

\section{Phase diagram obtained from Quantum Monte Carlo} 
\label{sec:MC}

Here we present results from a Quantum Monte Carlo study of the effective resonant subspace of the near-diagonally tilted decorated square lattice. We emphasize that we do not simulate the \emph{full} bosonic Hamiltonian: QMC would then look for the absolute ground state, which means that all bosons follow the tilt and fall down to minus infinity. Instead we simulate the effective resonant subspace, which is described by the constrain five state model.

\subsection{Order parameters\label{sub:Order-parameters}}

As we expect one-dimensional order
to persist in some region of the phase diagram, we introduce
order parameters which probe for 1D Ising order, as well as order parameters
which probe for two-dimensional order.

Before introducing our order parameters, we begin by reviewing the
order parameter of the one-dimensional system in Refs.~\onlinecite{SachdevMI,GreinerIsing}:
a staggered magnetization which breaks lattice symmetries (translation,
inversion, and reflection symmetry). In the language of the spin mapping
in Section~\ref{sub:spinmapping}, and Ref.~\onlinecite{GreinerIsing} it is given by
\[
M=\frac{1}{L}\sum_{l}(-1)^{l}\sigma_{l}^{z}.
\]
In the limit $\lambda_{y}=\infty$, $\lambda_{x}/\lambda_{y}=0$
each unit cell of the decorated square lattice has only three states
available, and we can directly translate this staggered magnetization
to our notation. Each chain aligned along the $x$ direction and at position
$y$ then has its own, independent order parameter
\begin{equation}
M_{{\rm LR}}(y)=\frac{1}{L_{x}}\sum_{x}\left(\hat{p}_{\rightarrow}-\hat{p}_{\leftarrow}\right)_{x,y}=\frac{1}{L_{x}}\sum_{x}m_{{\rm LR}}(x,y),\label{eq:LR-order-param}
\end{equation}
where $\hat{p}_{d}$ is a projection operator that projects onto the
dipole state $d.$
Note that this order parameter is normalized to take values in the interval $[-1,1]$.
The (staggered) magnetization of each unit cell
has been defined for dipoles along left-right direction only as
\[
m_{{\rm LR}}(x,y)=\hat{p}_{\rightarrow}-\hat{p}_{\leftarrow}
\]
When $\lambda_{y}\ne\infty,$ 
each unit cell has two additional
states available%
\footnote{This breaks the translation symmetry of the one-dimensional chain
explicitly: There is no translation relating the state $\ket{\rightarrow}$
to the state $\ket{\leftarrow}$, under which the order parameter
would change sign. There is, however, an inversion symmetry and a
reflection symmetry left, which can be spontaneously broken by an
ordered state. %
}. We can still use the above definition, and add another component
to the order parameter,
\[
m_{{\rm UD}}(x,y)=\hat{p}_{\downarrow}-\hat{p}_{\uparrow}
\]
the magnetization for dipoles aligned in y direction. We combine the
two to a vector,
\[
\vec{m}(x,y)=\left(\begin{array}{c}
m_{{\rm LR}}(x,y)\\
m_{{\rm UD}}(x,y)
\end{array}\right)
\]
we will refer to this as the ``magnetization'' of a unit cell of our
system. The magnetization $\vec{m}(x,y)$ transforms as a vector and is odd under inversion.
To probe for two-dimensional order in the system, we define the total
magnetization
\begin{equation}
\vec{M}=\frac{1}{L_{x}L_{y}}\sum_{x,y}\vec{m}(x,y).\label{eq:total-mag}
\end{equation}
and measure
$\ave{\vec{M}^{2}}$.
If the chains are aligned ferromagnetically, then this order parameer
is non-zero. To probe for antiferromagnetically aligned chains we
define a total staggered magnetization
\begin{eqnarray*}
\vec{M}_{{\rm stagg,x}} & = & \frac{1}{L_{x}L_{y}}\sum_{x,y}(-1)^{x}\vec{m}(x,y),\\
\vec{M}_{{\rm stagg,y}} & = & \frac{1}{L_{x}L_{y}}\sum_{x,y}(-1)^{y}\vec{m}(x,y).
\end{eqnarray*}

For an anisotropic tilt we expect the most important effect to be
an order within each chain. These chains may or may not be coupled
to form two-dimensional order. It is therefore usefull to define
order parameters which probe for one-dimensional order along $x$
or $y$ direction only. For this purpose we will use $M_{{\rm
LR}}(y)$, Eq.~\ref{eq:LR-order-param}, and average its square over all chains,
\begin{eqnarray}
\ave{\ave{M_{{\rm LR}}^{2}}}
& = & \frac{1}{L_{y}}\sum_{y}\ave{M_{{\rm LR}}^{2}(y)}
  = \frac{1}{L_{x}}\frac{1}{L_{y}^{2}}\sum_{x,x^{\prime},y}\ave{m_{{\rm LR}}(x,y)m_{{\rm LR}}(x^{\prime},y)}
 \label{eq:mag_1d}
\end{eqnarray}
and similarly for for chains aligned along $y$,
$ \ave{\ave{M_{{\rm UD}}^{2}}}= \frac{1}{L_{x}}\sum_{x}\ave{M_{{\rm UD}}^{2}(x)}.%
$


\subsection{Phase diagram}

\begin{figure}
\includegraphics[width=0.99\textwidth]{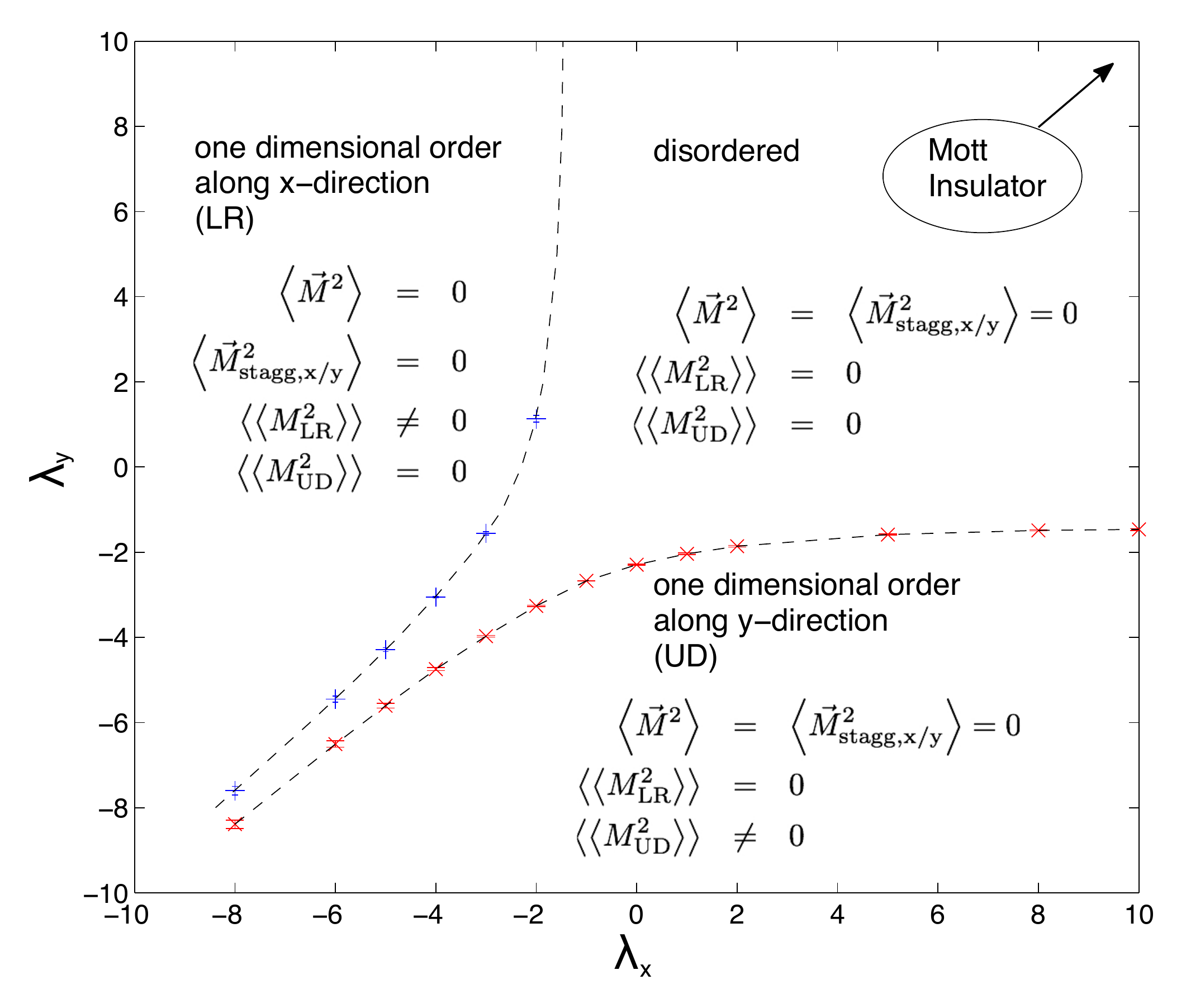}
\caption{\label{fig:DecSq-phase-diag}Phase diagram obtained from
QMC study. We find only one-dimensional order. The two-dimensional
order parameters are zero everywhere. The phase boundary was
obtained from crossing of the Binder\cite{binderCumulant} cumulant of the
one-dimensional magnetization, \eqnref{eq:mag_1d}, see Appendix \ref{appendix-qmc-binder}. Square lattice
of size $L_{x}=L_{y}=(4,8,16,32,64)$ the imaginary time slice
thickness was $a=0.04$, imaginary time direction was scaled with
the linear system size, $M_{\tau}=(40,80,160,320,640)$,
corresponding to temparatures
$T=(0.625,0.3125,0.1562,0.0781,0.0391).$ Essentially the same
phase diagram is obtained from order parameter scaling assuming
the Ising exponent $\eta=1/4.$ }
\end{figure}

Results of the QMC simulations are summarized in the phase diagram
shown in Figure~\ref{fig:DecSq-phase-diag}.
The disordered dipole state for
$\lambda_{x},\lambda_{y}\rightarrow-\infty$ appears to be
continuously connected to the parent Mott insulator at
$\lambda_{x},\lambda_{y}\rightarrow+\infty$ . There is a critical
line where the system undergoes a transition to an ordered state
with \emph{one dimensional} order along individual chains. For
$\lambda_{y}<\lambda_{x}$ these chains are aligned in $y$
direction (lower right corner of the phase diagram in
Fig.~\ref{fig:DecSq-phase-diag}), and $\ave{\ave{M_{{\rm
UD}}^{2}}}\ne0,$ while $\ave{\vec{M}{}^{2}}=\ave{\vec{M}_{{\rm
stagg,x}}^{2}}=\ave{\vec{M}_{{\rm stagg,y}}^{2}}=0.$ We will show
below that the system indeed seems to be disordered in the
transverse direction; 
each chain appears to have a two-fold degenerate ground state,
which is independenent of the order parameters of the neighboring
chains, and so the two-dimensional system has a ground state
degeneracy $2^{L_{x}}$, where $L_{x}$ is the linear system size in
$x$ direction, i.e. the number of chains. Correlations in the $x$
direction decay exponentially with a correlation length which is
smaller than the lattice spacing, $\xi_{x}\ll a$. There is no
region in the phase diagram where either $\ave{\vec{M}{}^{2}}$,
Eq.~\ref{eq:total-mag}, or $\ave{\vec{M^{2}}_{{\rm stagg,x}}}$ or
$\ave{\vec{M^{2}}_{{\rm stagg,}y}}$ takes a non-zero value.

\begin{figure}
\includegraphics[width=\textwidth]{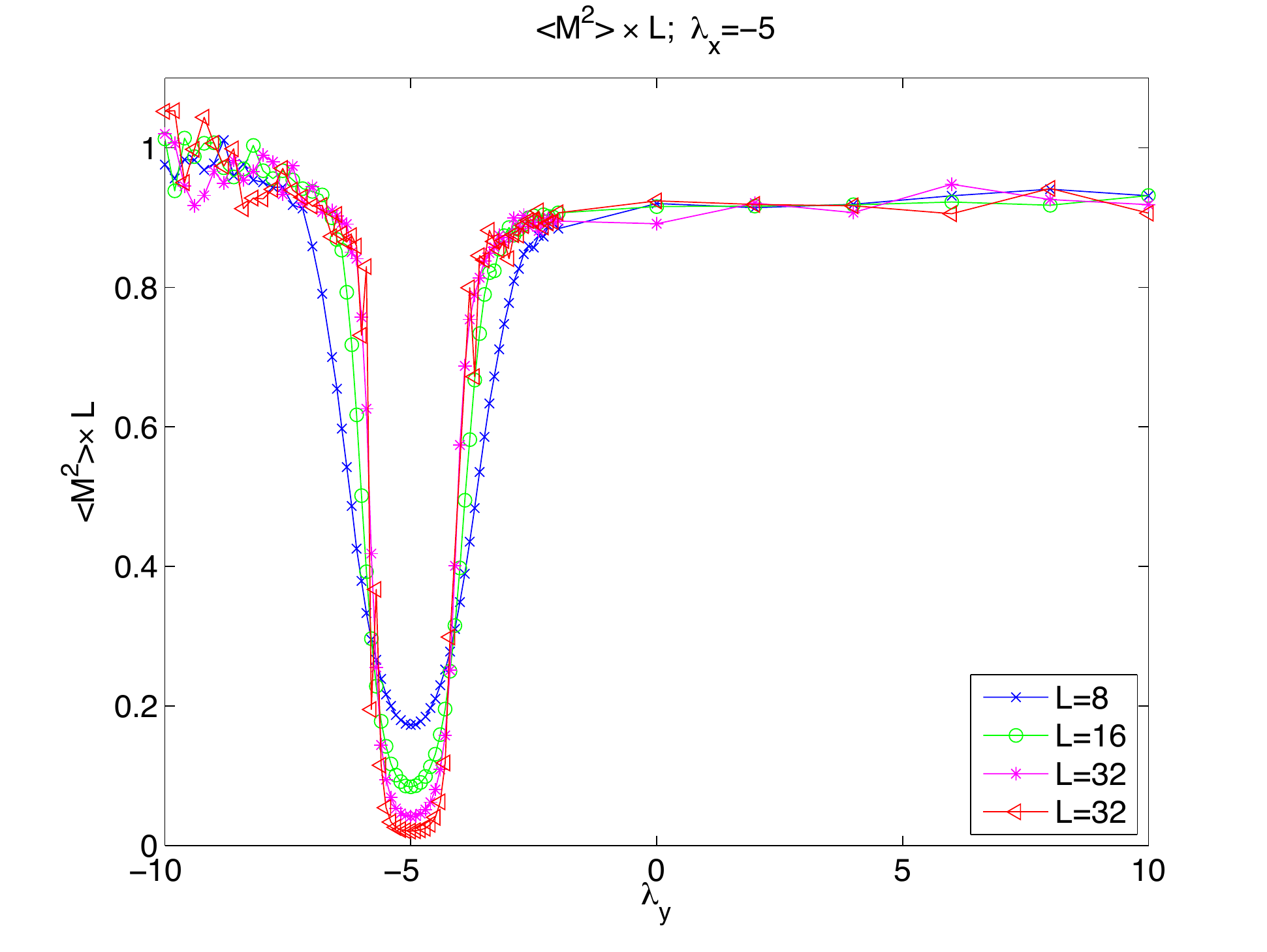}
\caption{\label{fig:twoDmag}
Absence of two dimensional order in the QMC study. Here we show the scaling of the total magnetization with system
size, $\ave{\vec{M}^{2}}\times L$. For randomly
aligned magnetized chains we expect $\ave{\vec{M}^{2}}\propto1/L$,
while for perfectly aligned chains $\ave{\vec{M}^{2}}={\rm const}$
and for antiferromagnetic order in transverse direction $\ave{\vec{M}^{2}}=0$.
In the disordered phase $\ave{\vec{M}^{2}}\propto1/(L_{x}L_{y})$.
This shows that in the ordered phase the system behaves as a
collection of independent chains with one-dimensional order within each chain. Here $\lambda_x=-5$ was kept fixed; similar
results are obtained for different cuts through the phase diagram. }

\end{figure}

We probe for two-dimensional order by measuring $\ave{\vec{M}^{2}}$.
This quantity scales with system size exactly as it would for a collection \emph{randomly}
aligned magnetized one-dimensional chains. This suggests that there
is no two-dimensional order: while in some regions of the phase diagram
there is long range order along individual chains,
there is no correlation between these chains,
see Figure~\ref{fig:twoDmag}.

We find good data collapse with the critical exponents of the 2D classical Ising model (Onsager exponents), see
Appendix. 
This further supports the observation that individual chains behave as independent 1d Ising systems.

Note that this QMC method can also be used for a ``doped'' system, i.e. a system with initial defects in the Mott insulator. Defected sites then block neighboring links from dipole occupation. Studying such a disordered system is an interesting subject to future work. 

\section{Inter chain coupling: Exact diagonalization study}
\label{sec:inter-chain-coupling}

Quantum Monte Carlo results describe a system which decouples into a collection
of one-dimensional chains, while 
on symmetry grounds 
one would expect that a coupling between the chains should be
generated\cite{SubirBook}. The following scenarios might explain this disagreement
\begin{itemize}
\item \emph{hidden symmetry}: there could be a subtle microscopic
symmetry which forbids any coupling between the order parameters
of the chains. The next relevant term is then the energy-energy
coupling, which leads to a change of the critical exponent $\nu$
but no ordering of the chains; or \item \emph{finite-size
effects:} the coupling between the chains could be present, but is
too small to 
have an observable effect for the system sizes studied with QMC.
Simulating larger lattices should then, in principle, find a phase
with two-dimensional order.
\end{itemize}
Here we resolve this question
by an exact diagonalization study of a toy
model, which consists of two chains and qualitatively captures the
interchain constraint, and thus the coupling.

We find that a coupling between two neighboring chains is indeed
present: it is antiferromagnetic in sign and very small in
magnitude. It appears only in high order in perturbation theory in
$t/|U-E|$
, and is thus not observable for
realistic system sizes, neither in QMC nor in cold-atomic quantum
simulation experiments.

\subsection{Toy model}

\begin{figure}[tbh]
\begin{center}
\includegraphics[width=0.6\textwidth]{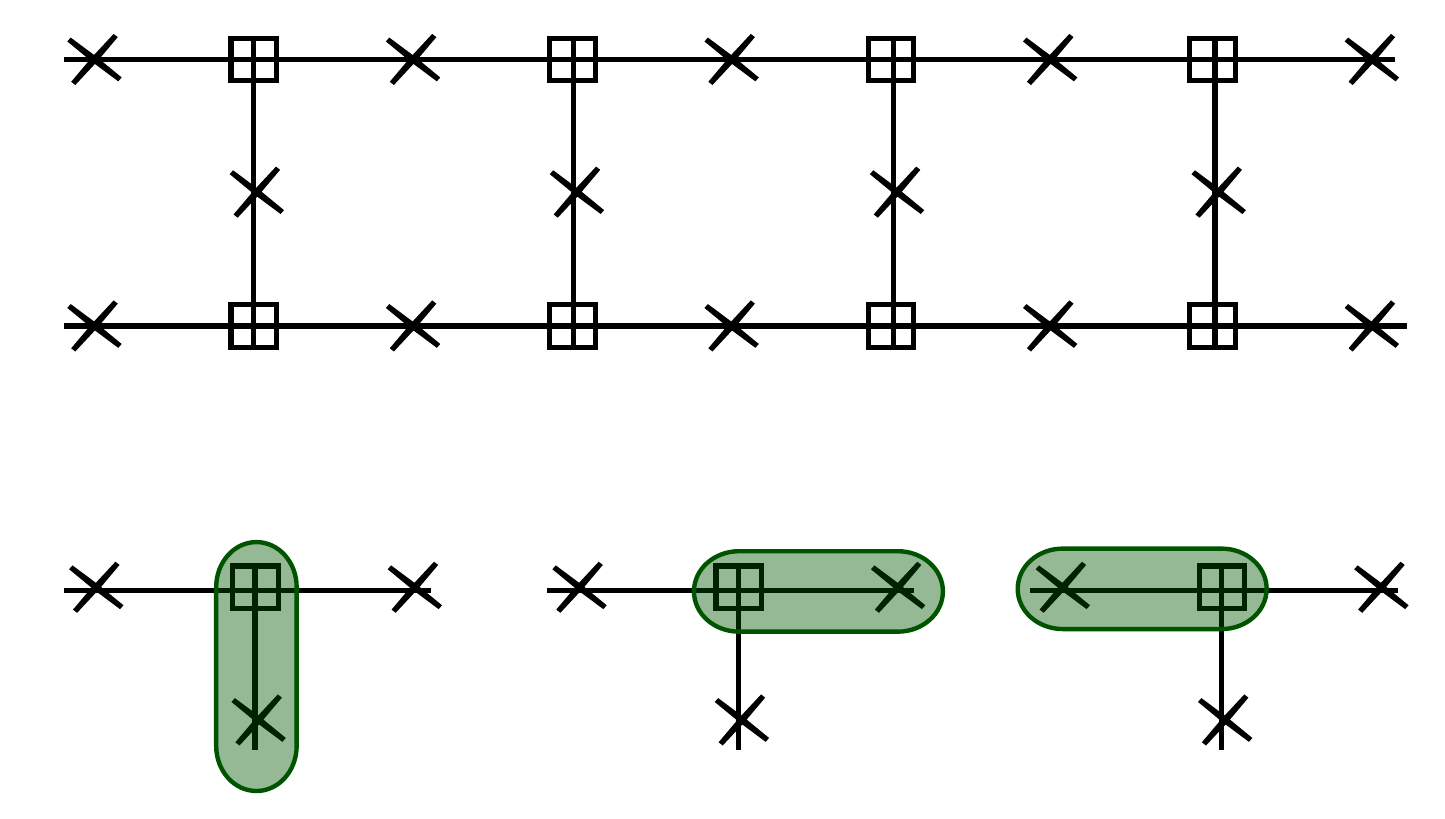}
\end{center}
\caption{\label{fig:Two-chain-model}Two chain model used for the exact diagonalization study. Each central site (marked
with a square) can be in one out of three states. A chain of length
four unit cells is shown. This model should capture the interchain behavior of our
model qualitatively. }
\end{figure}

We use a simplified model 
with only two chains. We reduce the Hilbert space further by only keeping 
the three most relevant dipole states in each unit cell\footnote{'unit cell' here 
refers to the unit cell of each individual chain, not the the unit cell of the combined ladder.}, see Figure~\ref{fig:Two-chain-model}. The state without dipole has been integrated out\footnote{this of course only works for $\Delta_{x},\Delta_{y}<0$%
}, while one of the 
vertical dipole states is missing: this enhances the coupling and
reduces the size of the Hilbert space, which enables us to study
longer chains. We expect this toy model to qualitatively describe
the interchain coupling of our system.

The Hamiltonian for each of a single unit cell reads 
\[
H_{{\rm site}}=\left(\begin{array}{ccc}
\Delta_{b} & -t_{b} & -t_{b}\\
-t_{b} & \Delta_{a} & -t_{a}\\
-t_{b} & -t_{a} & \Delta_{a}
\end{array}\right)
\]
where $\Delta_{a}$ is the energy cost for a dipole along chain direction,
and $\Delta_{b}$ is the energy cost for a dipole in direction transverse
to the chains. The effective hopping elements $t_{a},$ and $t_{b}$
are obtained from second order perturbation theory, and they both
have a negative sign (since $\Delta_{a}<\Delta_{b}<0$):
\begin{eqnarray}
t_{b} & = & t^2\left(\frac{1}{\Delta_{a}}+\frac{1}{\Delta_{b}}\right),\\
t_{a} & = & \frac{2t^{2}}{\Delta_{a}}.
\end{eqnarray}
In addition
to the single site Hamiltonian, there is the hard-core constraint
forbidding two central sites from pointing toward each other. This
constraint reduces the size of the Hilbert space. 

\subsection{Results}

We diagonalized this system for a chains of up to length
nine\footnote{ while we can diagonalize chains of length nine in
all regions of the phase diagram, in some regions the splitting
between the lowest eigenvalues seems to become smaller than
machine precision, this limits our analysis.}, with periodic
boundary conditions. The chain length $L$ refers to the number of unit cells in each chain. 
In analyzing the results, it is useful to
compare the spectrum to that of a simple model of two decoupled
Ising chains in the ordered phase. For systems of finite length,
this model would give four low-energy states, whose splitting
vanishes exponentially with the system size in the thermodynamic
limit.

\begin{figure}[tbph]
\begin{center}
\includegraphics[width=0.45\textwidth]{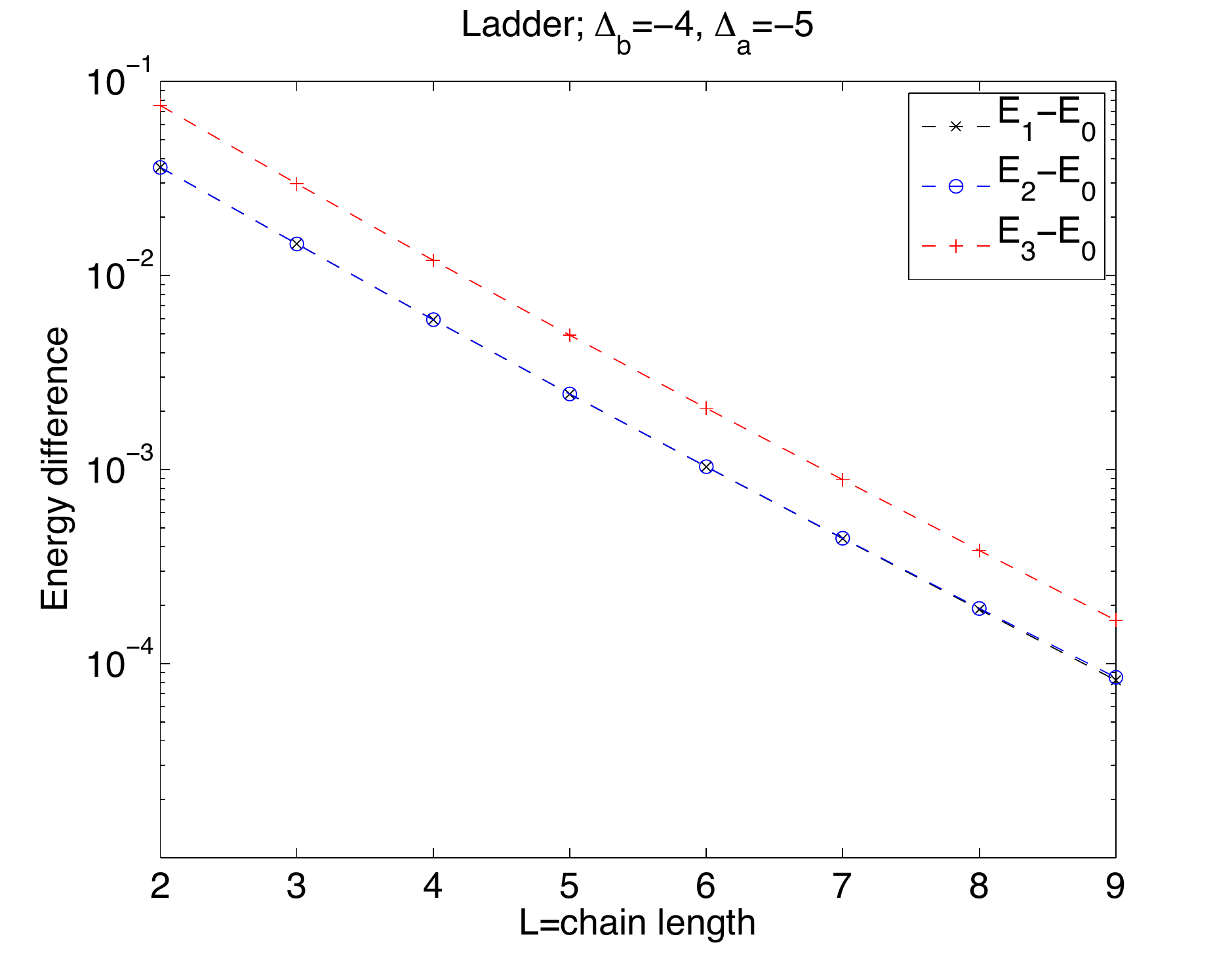}
\includegraphics[width=0.45\textwidth]{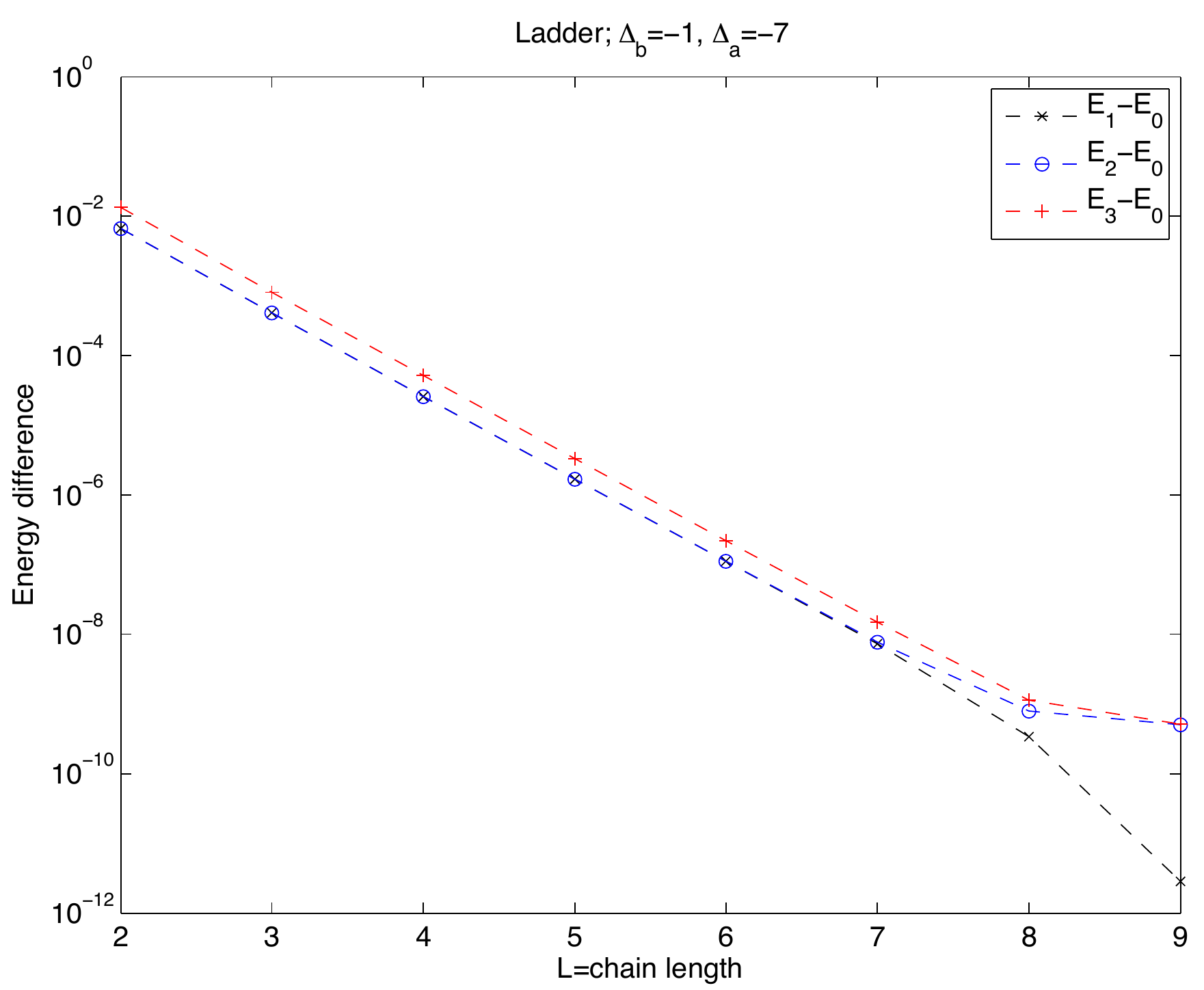}
\end{center}
\caption{\label{fig:four-gs-in-ordered-phase}
Strong finite-size effects:
Logarithmic plot for energy of
the first three excited states relative to the ground state as a function
of system length, for (a) $\Delta_a=-5, \Delta_b=-4$ (b) $\Delta_a=-7, \Delta_b=-1$.
The lines $E_{1}-E_{0}$ and $E_{2}-E_{0}$ appear on top of
each other. At first sight this suggests that there are four ground states in the thermodynamic limit and that thus the order parameters of the chains
are not coupled. However there is a small splitting between the first two levels above the ground state.
As can be seen in Fig.~\ref{fig:Splitting-E1-E2} this splitting grows linearly with system size and eventually dominates over the finite-size gap for long enough chains.}
\end{figure}

Fig. \ref{fig:four-gs-in-ordered-phase} shows the energy of the
three lowest excited states relative to the ground state, as a
function of the system size.
At first glance, these results are consistent with the model of completely decoupled chains, 
since the energy splitting to the three lowest excited states
appears to vanish exponentially. However, for certain parameters
and for the longest systems (Fig.
\ref{fig:four-gs-in-ordered-phase}b), we observe a deviation from
the decoupled chain model. Two of the low-energy states remain
nearly degenerate, while the gap to the other two starts deviating
from exponential. This behavior is consistent with having a small
but non-zero inter-chain coupling.

\subsubsection{Estimate of the order parameter coupling}

The following observations further support the existence of a
small coupling between the order parameters of the two chains:
\begin{enumerate}
\item There is a splitting between the first and the second
excited state, $E_{2}-E_{1}$. This splitting is too small to be
visible in Fig. \ref{fig:four-gs-in-ordered-phase}a. We present it
in Fig.~\ref{fig:Splitting-E1-E2}. Although the magnitude of the
 splitting is very small, it grows approximately linearly
with system size, consistent with a finite energy density
associated with an inter-chain order parameter coupling. 
\\
\begin{figure}
\begin{center}
\includegraphics[width=0.8\textwidth]{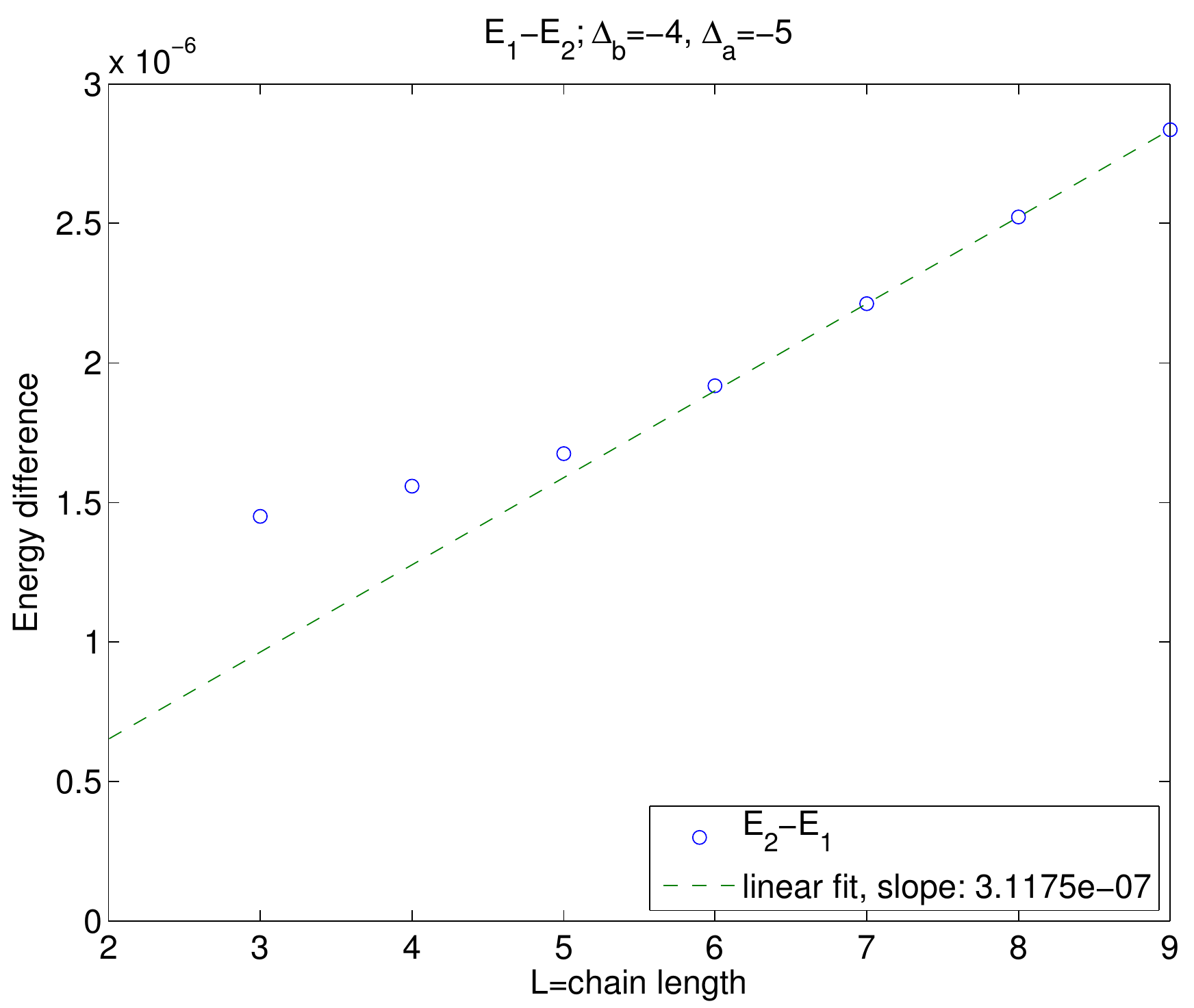}
\end{center}
\caption{\label{fig:Splitting-E1-E2} Splitting between first and
second excited state grows linearly with system size. This
suggests that there is indeed a coupling between the order
parameters of these two chain.}
\end{figure}

\item We can show that there is no hidden symmetry forbidding order parameter coupling:
if we fix the boundary conditions%
\footnote{We can view the segment under consideration as a part of a very long
system, which is aligned either ferromagnetically or antiferromagnetically.
The remainder of the long system provides the boundary conditions
for the segment. %
} to make the chains either aligned or anti-aligned, there is a
difference in ground state energy of these two systems which grows
linearly with system size, see Fig.~ \ref{fig:boundary-cond}. We
can use the slope of this curve as an estimate for the coupling of
the order
parameters per unit cell. \\
\begin{figure}
\begin{minipage}{0.3\textwidth}
\includegraphics[width=\textwidth]{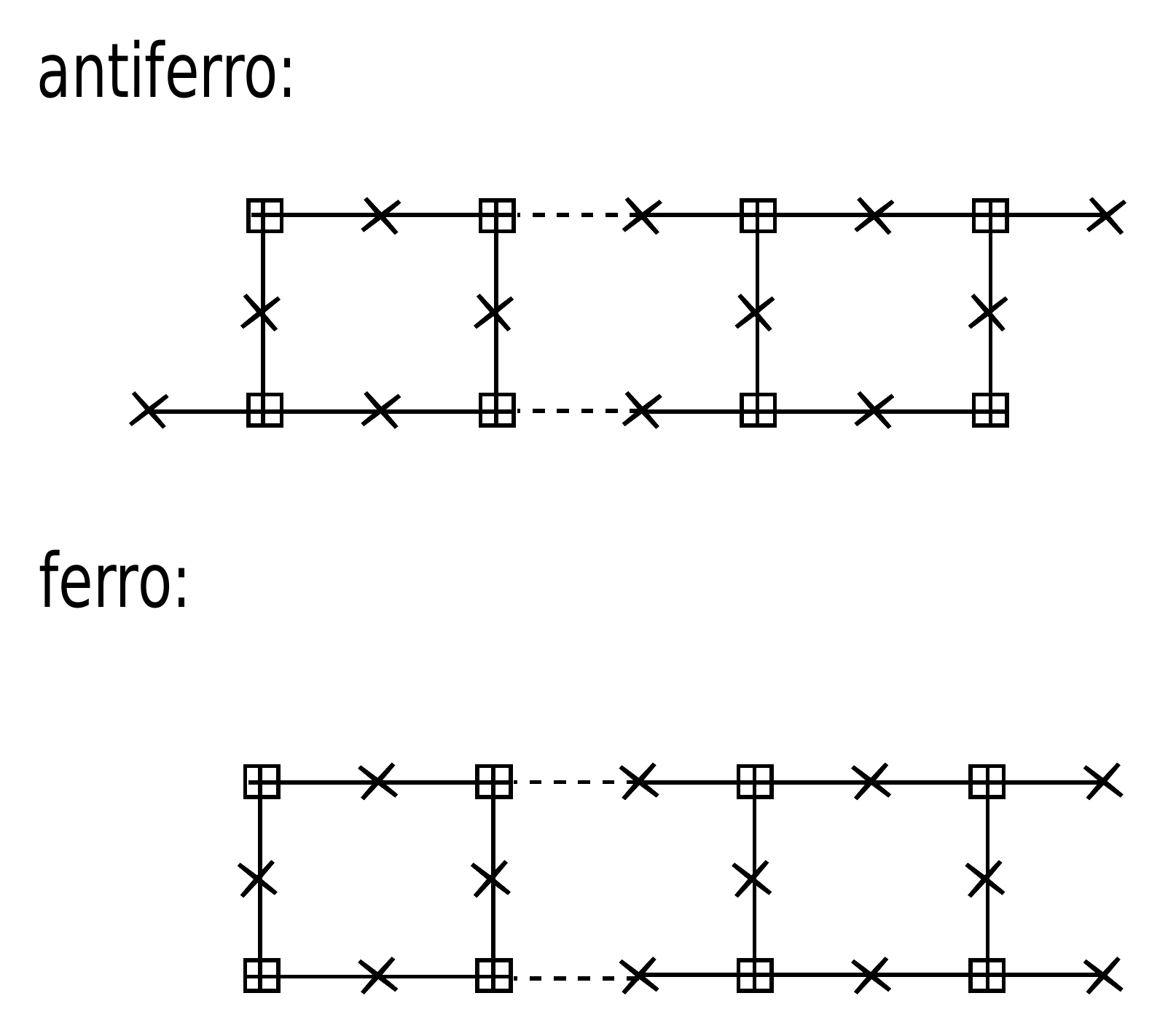}
\end{minipage}
\begin{minipage}{0.68\textwidth}
\includegraphics[width=\textwidth]{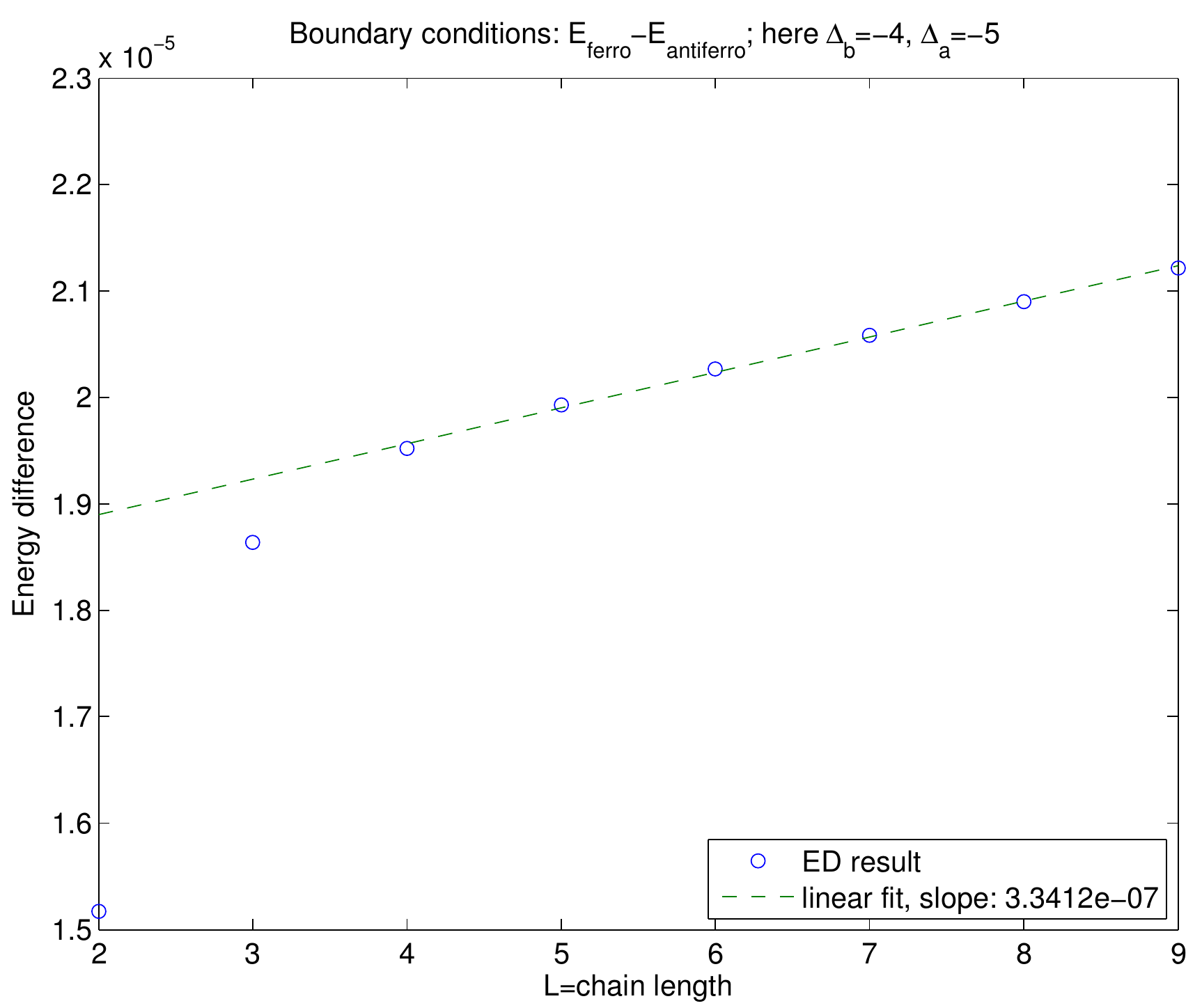}
\end{minipage}
\caption{ \label{fig:boundary-cond} The inter-chain coupling is
antiferromagnetic, which we show by fixing the boundary conditions
so that the chains are forced to be either aligned or
anti-aligned. The energy difference grows linearly with system
size and antiferromagnetic boundary conditions are energetically
favorable. }
\end{figure}

\end{enumerate}
The slopes of both curves in Fig.
\ref{fig:four-gs-in-ordered-phase},\ref{fig:boundary-cond} agree
approximately, giving us a good estimate for the order parameter
coupling of the chains. For the system sizes we studied, 
however, this coupling is smaller than the effective tunneling
element between the different ground states, i.e. smaller than the
finite-size gap. For larger chains this order parameter coupling
will start to dominate over the tunneling, and then there will
only be two ground states. The coupling between two chains is
\emph{antiferromagnetic.}

\subsubsection{Estimate of order in perturbation theory for coupling}
\begin{figure}
\begin{center}
\includegraphics[width=0.8\textwidth]{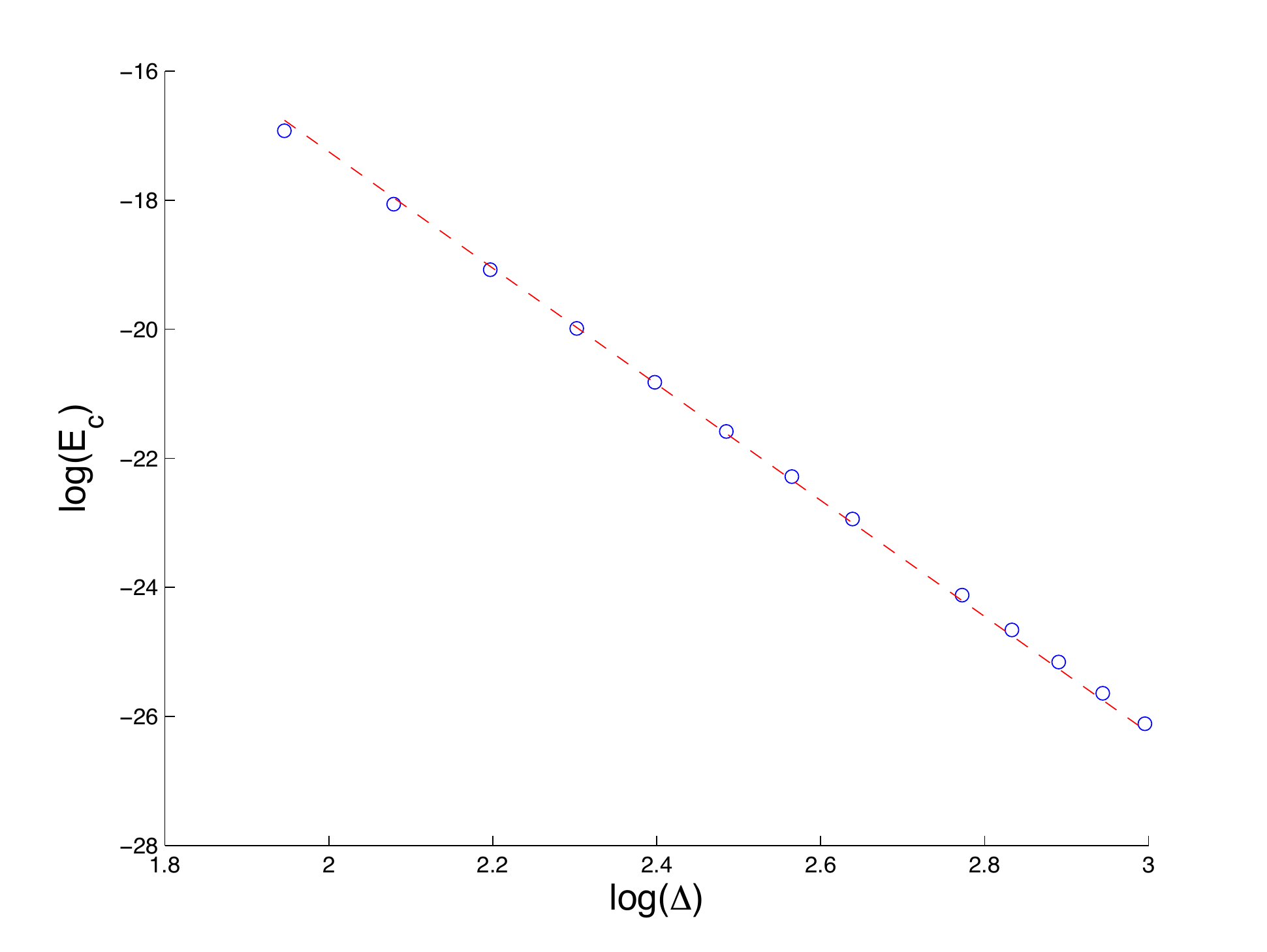}
\end{center}
\caption{\label{fig:order-perturbation}
The antiferromagnetic coupling appears in 10th order in perturbation theory. This logarithmic plot shows that the coupling per unit length scales as $\abs{\Delta}^{-9}$, where $\Delta=\Delta_b-\Delta_a$.
}
\end{figure}

We now estimate the order in perturbation theory in which the
antiferromagnetic coupling is generated. To this end, we set
$t_a=t_b=1$ and use $\Delta=\Delta_b-\Delta_a$ as our only tuning
parameter. Let $E_c$ be the energy per unit length of the
antiferromagnetic coupling
$$
E_c=\frac{E_2-E_1}{N}
$$
(here $N$ is the chain length). If the antiferromagnetic coupling appears roughly in $n$th order in perturbation theory, then
\begin{eqnarray}
E_c&\propto&\abs{\Delta}^{1-n}\\
\log\left(E_c\right)&=&(1-n) \log\left(\abs{\Delta}\right)+\rm{const}
\end{eqnarray}
We plot $\log{E_c}$ versus $\log(\abs\Delta)$ and obtain $n=10$
from a linear fit (see Fig. \ref{fig:order-perturbation}). This
suggests that the antiferromagnetic inter-chain coupling is
generated roughly in 10th order in perturbation theory within this toy
model, and in 20th order in perturbation theory within the bosonic
model.

\subsection{Mechanism for inter-chain coupling}

Having established that, in the ordered phase, the chains couple
antiferromagnetically, it is natural to ask what is the
microscopic mechanism responsible for this coupling. Below, we
give a qualitative argument for the generation of inter-chain
coupling in high orders in $t/\Delta$, which predicts that the
sign of the coupling should be antiferromagnetic.

\begin{figure}
\begin{center}
\includegraphics[width=\textwidth]{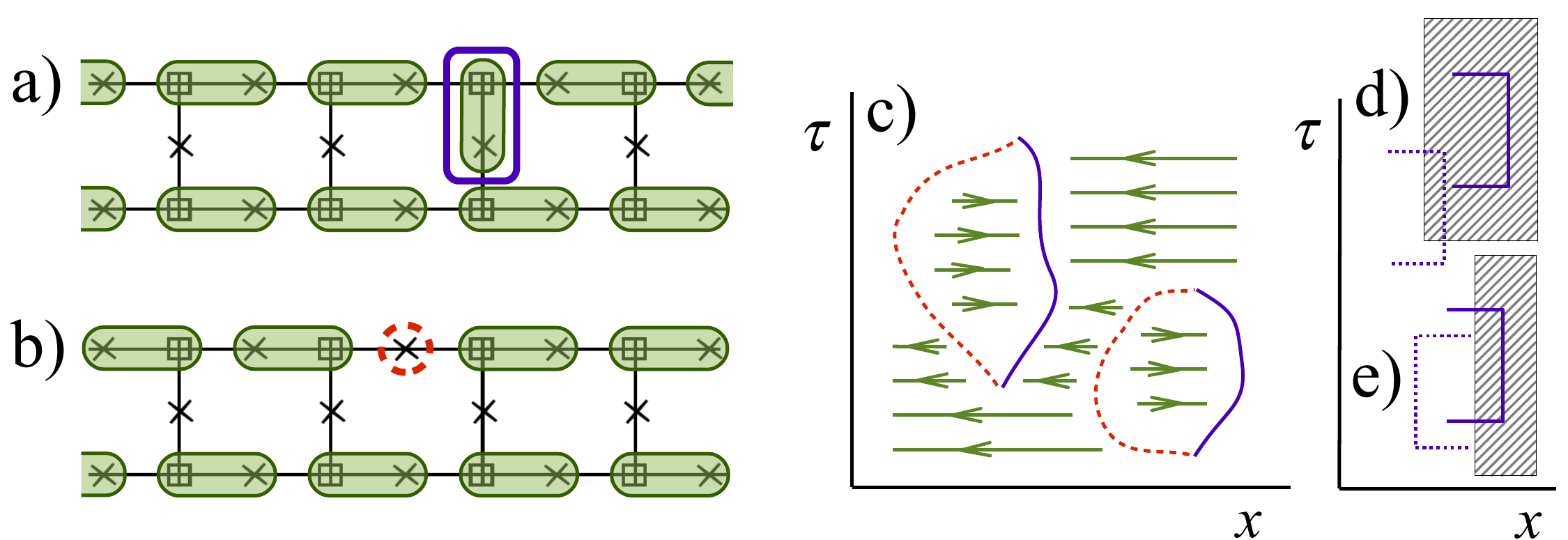}
\end{center}
\caption{\label{fig:kinks} A kink (a) and anti-kink (b)
configuration. Kinks in different chains cannot reside on the same
rung, because of the hard-core constraint; no such restriction
exists for the anti-kinks. c) Schematic space-time trajectories
for kink-anti-kink pairs. The kinks (anti-kinks) are represented
by solid (dashed) lines, respectively. The arrows represent the
direction of the Ising order parameter. The average order
parameter is assumed to be pointing to the left. In this
configuration, the kinks are typically to the \emph{right} of the
anti-kinks, and their world lines typically curve \emph{left}.
Since the two chains couple only through the kinks, this provides
a mechanism for coupling the order parameters of the two chains.
d,e) A simplified model showing the source of the interaction
between the order parameters of the two chains.
The space-time trajectories of kinks
on the upper chain are approximated, for simplicity, by the
rectangular U-shaped curves shown in blue. We define the ``position"
of a world line as the center of the rectangle enclosing it. The
hatched region
shows the space-time volume blocked for the position of kinks in the
lower chain, assuming that the order parameters
of the two chains are aligned (d) or anti-aligned (e), due to the hard
core repulsion between kinks in the upper and lower chains. The
space-time trajectories of kinks in the lower chain in the two cases
are shown by the dashed blue curves.}
\end{figure}

We begin by the observation that in the Ising ordered phase, there
are two distinct types of domain wall fluctuations, a kink and an
anti-kink (see Fig. \ref{fig:kinks}(a,b)). Kink-anti-kink pairs
can be generated virtually, lowering the kinetic energy. We note
also that the two domain walls have a parametrically different
effective mass: the kink (Fig. \ref{fig:kinks}a) can hop via a
process of order $t_b^2/(|\Delta_a-\Delta_b|)$, while the
anti-kink (Fig. \ref{fig:kinks}b) is much lighter, hopping via a
process of order $t_a$. Therefore, the anti-kinks are more
delocalized.
We note also that kink-anti-kink pairs in a fully ordered
configuration are created with a preferred orientation. For
example, if the order parameter in a particular chain is pointing
to the left, then each kink is typically to the right of its
anti-kink partner (see Fig~\ref{fig:kinks}c). 
As a consequence, world lines
of kinks curve in the opposite direction than world lines of
anti-kinks; and which way they curve is determined by the sign of
the order parameter.
Some typical space-time paths of fluctuating
kink-anti-kink pairs are shown schematically in Fig.
\ref{fig:kinks}c.

The order parameters of two neighboring chains are not coupled
directly; i.e., in a classical, fully ordered configuration (which
is the ground state in the limit $t_{a,b}=0$), there is no energy
difference between an aligned and an anti-aligned configuration.
However, in the presence of quantum fluctuations, the two chains
become coupled. For instance, in our two-chain model, kinks cannot
occur simultaneously on both chains, because of the constraint
preventing two dipoles to point towards each other. Note that
there is no such constraint for anti-kinks; configurations in
which anti-kinks in the two chains occur on the same rung are
allowed.

This hard-core interaction between kinks provides a mechanism for
coupling the order parameters of the two chains. Consider the kink
space-time configuration in Fig. \ref{fig:kinks}c.
In this configuration the world lines of kinks tend to curve to
the right, because of the hard-core repulsion between kinks and
anti-kinks 
on the same chain. 
The repulsive interaction between kinks on 
two 
neighboring
chains reduces
the ``phase space'' available for quantum fluctuations. However,
one can imagine that if the two order parameters are anti-aligned,
the phase space for fluctuations is slightly \emph{larger} than in
the opposite case. This is since if the order parameters are
anti-aligned, the space-time trajectories of the kinks on the
lower chain tend to curve oppositely to those of on the upper
chain, allowing one to fit more quantum fluctuations in a given
space-time ``volume'' (thus lowering the kinetic energy). To see
this, imagine for simplicity that the space-time trajectories of
kinks in the upper chain are of the shape shown by the solid blue
curve in Fig. \ref{fig:kinks}d. Then, one can ask how much
space-time volume does a kink in the upper chain block for kinks
in the lower chain. The blocked volume is twice as large in the
case in which the order parameters of the two chains are aligned,
compared to the anti-aligned case (Fig. \ref{fig:kinks}d,e),
favoring the order parameters to be anti-aligned. This ``order by
disorder'' mechanism explains how the repulsion between kinks can
generate a coupling between the order parameters of the two chains
to high order in $t_{a,b}/|\Delta_a-\Delta_b|$. Moreover, it
predicts that the coupling is \emph{antiferromagnetic},
consistently with the ED findings described above.
We believe that it is unlikely that
this coupling could introduce new phases.
\section{Conclusions}
\label{sec:Conclusions}

We have proposed a setup to simulate frustrated quantum Ising
spins with cold atoms 
in a tilted optical lattice, by generalizing an idea which has
been successfully applied experimentally in one dimension. We have
studied the phase diagram of the resulting model and found that it
has strong finite-size effects. For realistic systems it
decouples into a collection of one dimensional Ising chains, a
coupling between the chains is present in the thermodynamic limit.
A quantum simulator of ultracold atoms would, however, be limited
in the system size. We therefore expect it to observe a
one-dimensional transition to ordered chains, just as we did in
QMC.

\acknowledgements

We thank M.~Greiner and J. Simon for useful discussions. This
research was supported by the National Science Foundation under
grants DMR-1103860, DMR-0757145 and DMR-0705472, and by a MURI
grant from AFOSR.
S.P. acknowledges support from the Minerva Stiftung.

\appendix

\section{Details on the QMC study}
\label{appendix-qmc}

We mapped the two-dimensional quantum model \eqnref{eq:dipole-model} to three-dimensional classical model with discrete imaginary time. The model has now become a three-dimensional constraint five-state model on a cubic lattice. In the Monte Carlo simulations we used single site flips (which change the state of a unit cell), as well as cluster updates (i.e. updates which  flip a segment in imaginary time at a given spatial position, and updates that flip the magnetization of an entire chain). 
Time discritization introduces a Trotter error, but it should change neither the nature of the phases, nor the universality class of the phase transition. 

For the results presented in this paper we used an imaginary time slice thickness $a=0.04$; we have also run simulations with smaller imaginary time slices and obtained similar results. We had 27000 equilibration sweeps before starting to take measurements. Mearsurements were taken every 50 sweeps, and 50 measurements were binned into one group. The number of groups were (2290, 1090, 530, 210) for linear system size (8, 16, 32, 64). Error bars were obtained from a simple binning analysis\cite{BinderMCbook}.

\subsection{Binder cumulant }
\label{appendix-qmc-binder}
\begin{figure}[tbp]
\begin{center}
\includegraphics[width=0.8\textwidth]{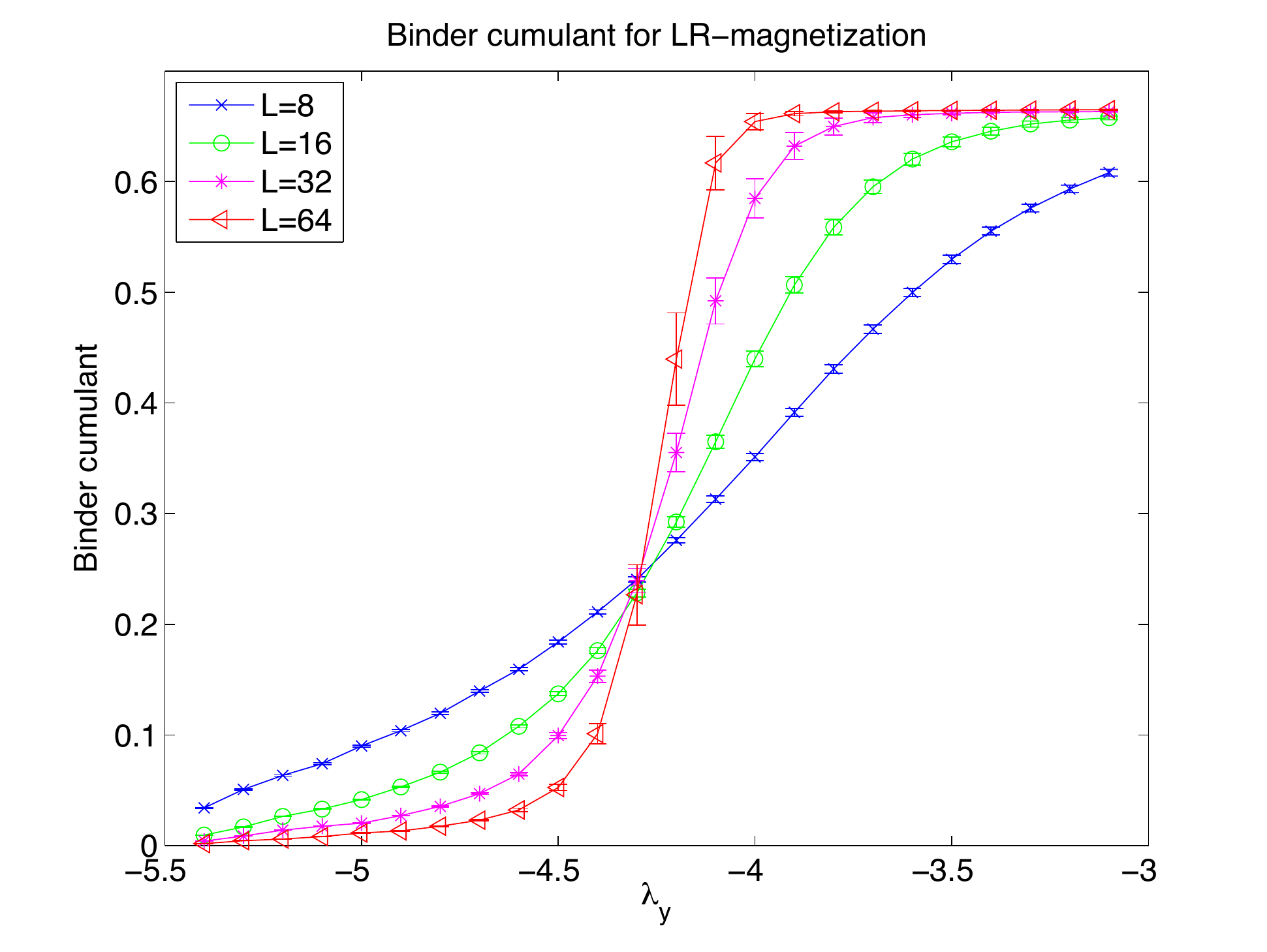}\\
\includegraphics[width=0.8\textwidth]{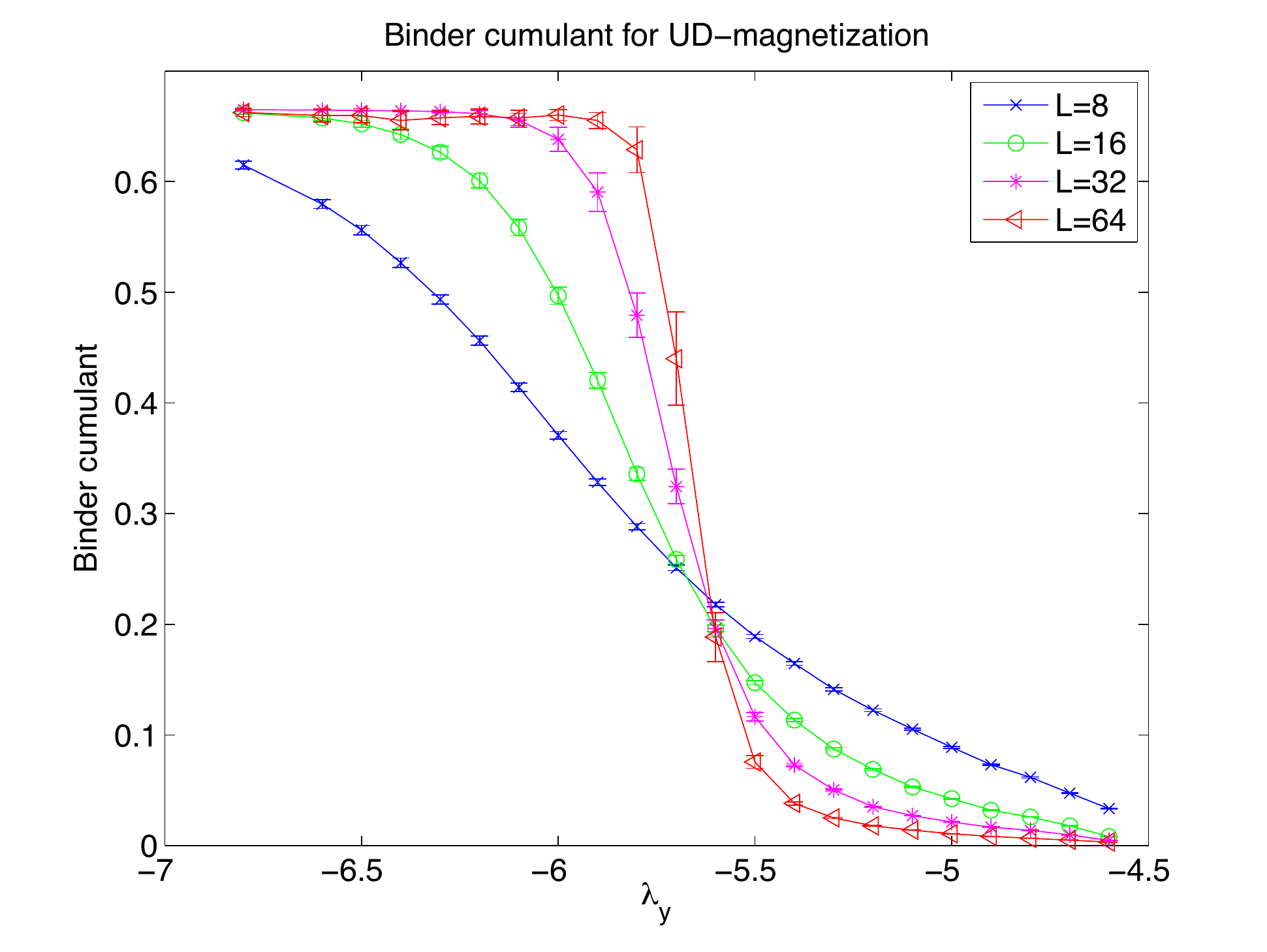}
\end{center}
\caption{\label{fig:Binder-cumulant}Binder cumulant for order parameters $\ave{\ave{M_{{\rm LR}}^{2}}}$
and $\ave{\ave{M_{{\rm UD}}^{2}}}$ for a horizontal cut through the
phase diagram, keeping $\lambda_{x}=-5$ fixed, plotted for different
system sizes, $L_{x}=L_{y}=8,\:16,\:32,\:64.$ Similar plots are obtained for different cuts through the phase diagram.}
\end{figure}

The Binder cumulant gives a good estimate for the critical point,
and it does not depend on critical exponents~\cite{binderCumulant}.
For an Ising order parameter, $M$, the Binder cumulant is defined
as
\[
U=1-\frac{\left\langle M^{4}\right\rangle }{3\left\langle M^{2}\right\rangle ^{2}}.
\]
When the Binder cumulant is plotted for different system sizes, all
curves should cross at the critical point for the following reason.
While for an infinite system the magnetization vanishes at the critical
point as
\[
M\propto\left(-\tau\right)^{\beta},
\]
where $\tau$ is the reduced temperature $\tau=T/T_{c}-1$, $\beta$ is the critical exponent of the
magnetization. For
a finite system there are corrections to scaling, described by a scaling
function $\phi$, which only depends on $\xi/L$,
\[
M=\left(-\tau\right)^{\beta}\phi\left(\xi/L\right)=\left(-\tau\right)^{\beta}\tilde{\phi}\left(\tau L^{1/\nu}\right),
\]
where $\xi$ is the correlation length, and $\nu$ is the correlation length exponent.
We used $\xi=\tau^{-\nu}$ to rewrite the scaling function for a different
argument. The average magnetization squared, and raised to the fourth
power, have different scaling functions,
\begin{eqnarray*}
\left\langle M^{2}\right\rangle  & = & \left(-\tau\right){}^{2\beta}u_{2}\left(\tau L^{1/\nu}\right)\\
\left\langle M^{4}\right\rangle  & = & \left(-\tau\right){}^{4\beta}u_{4}\left(\tau L^{1/\nu}\right)\\
\end{eqnarray*}
and so the Binder cumulant is a function of this same argument, $\tau L^{1/\nu}$
\begin{equation}
U(\tau,L)=1-\frac{u_{4}\left(\tau L^{1/\nu}\right)}{3\left(u_{2}\left(\tau L^{1/\nu}\right)\right)^{2}}=f\left(\tau L^{1/\nu}\right),\label{eq:binder-scaling}
\end{equation}
At the critical point we have $\tau=0$, and so the Binder cumulant
at this point should not depend on system size.
In the thermodynamic limit for an Ising system in the ordered phase
$U\rightarrow\frac{2}{3}$, and $U\rightarrow0$ in the disordered
phase. Fig.~\ref{fig:Binder-cumulant} shows the Binder cumulant
of the one-dimensional order parameters $\ave{\ave{M_{LR}^{2}}}$
and $\ave{\ave{M_{UD}^{2}}}$ for different cuts through the phase
diagram.

\subsection{Order parameter scaling and the critical exponent $\eta$}

\begin{figure}
\begin{center}
\includegraphics[width=0.8\textwidth]{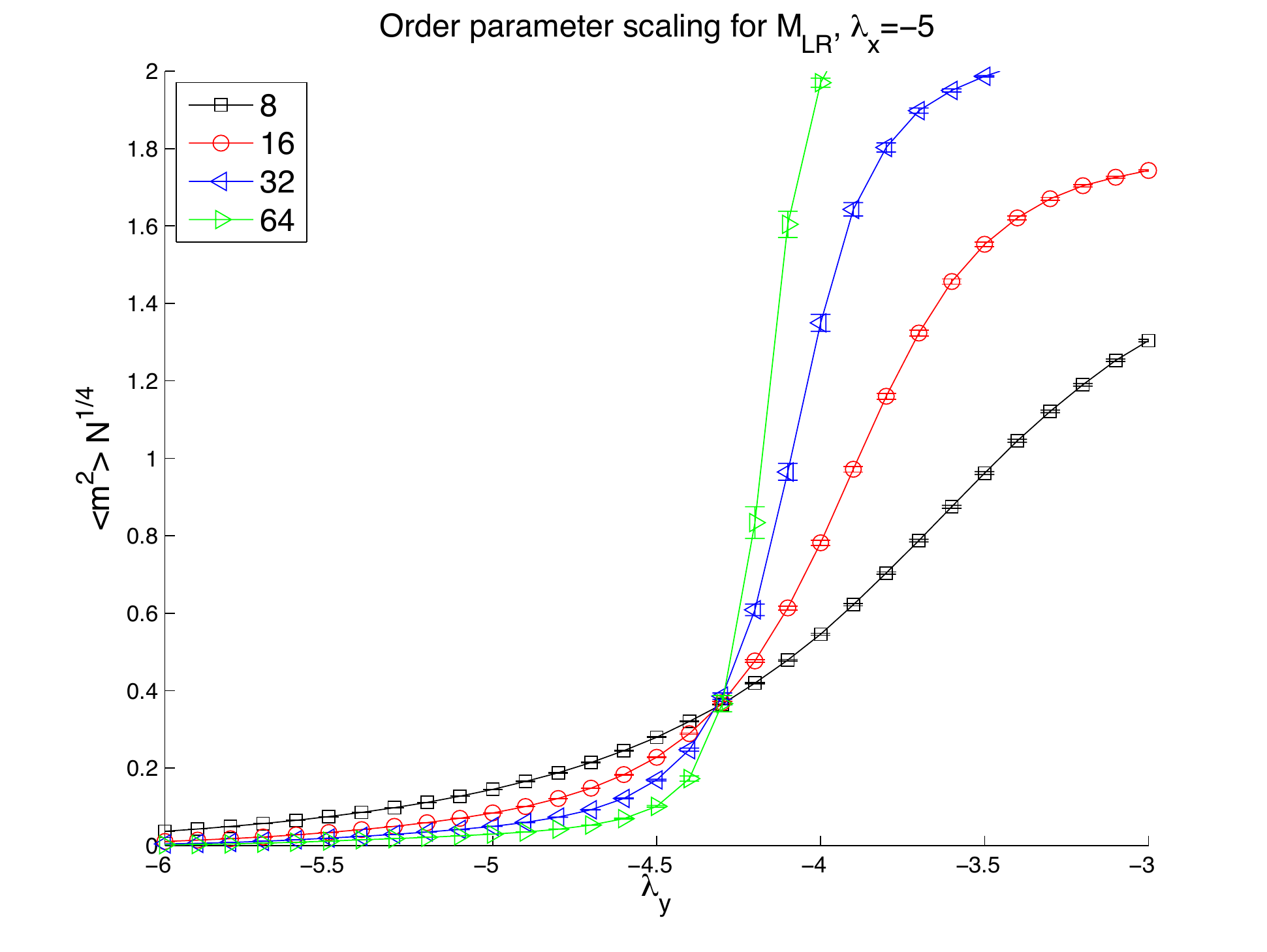}\\
\includegraphics[width=0.8\textwidth]{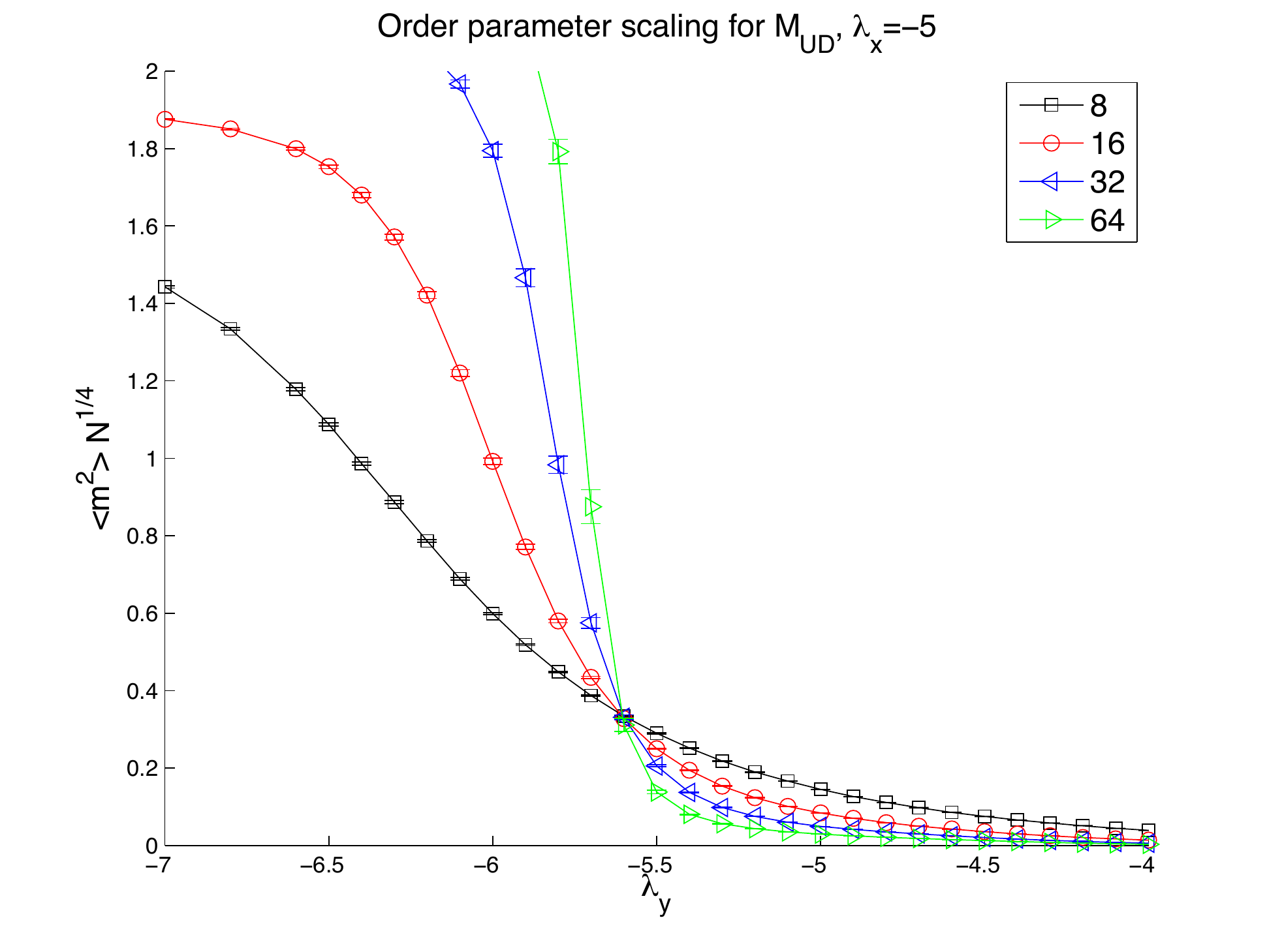}
\end{center}
\caption{\label{fig:Order-paramter-scaling}Order parameter scaling, assuming
the correlation length exponent $\eta=1/4$ of the classical two-dimensional
Ising model. Good agreement of the crossing points, also with the
ones found from the Binder cumulant. }
\end{figure}

The phase transition point can also be found from order paramter scaling.
This will depend on the correlation length exponent, $\eta$. Let
$M$ again be an Ising order parameter, at the critical point the
correlation decays as a power law,
\begin{eqnarray}
\ave{M(r)M(0)} & \propto & 
\abs{\vec{r}}^{-(d+z-2+\eta)}
\end{eqnarray}
$z$ is the dynamic critical exponent, which is $z=1$ in case of the Ising model. We define
M as the (normalized) total magnetization $M=1/L\int M(r) dr$, and for a one dimensional system ($d=1$) we have
\begin{equation}
\ave{M^{2}} \propto  L^{-\eta}
\end{equation}
The correlation length exponent for the 2D classical Ising
model is $\eta=1/4.$ In Figure~\ref{fig:Order-paramter-scaling}
we plot $\ave{\ave{M_{LR}^{2}}}L^{-1/4}$ and $\ave{\ave{M_{UD}^{2}}}L^{-1/4}$
for different system sizes. The crossing point of these lines gives us the phase boundary which agrees
with the one found from the Binder cumulant.

\subsection{Critical exponent $\nu$}

\begin{figure}
\begin{center}
\includegraphics[width=0.8\textwidth]{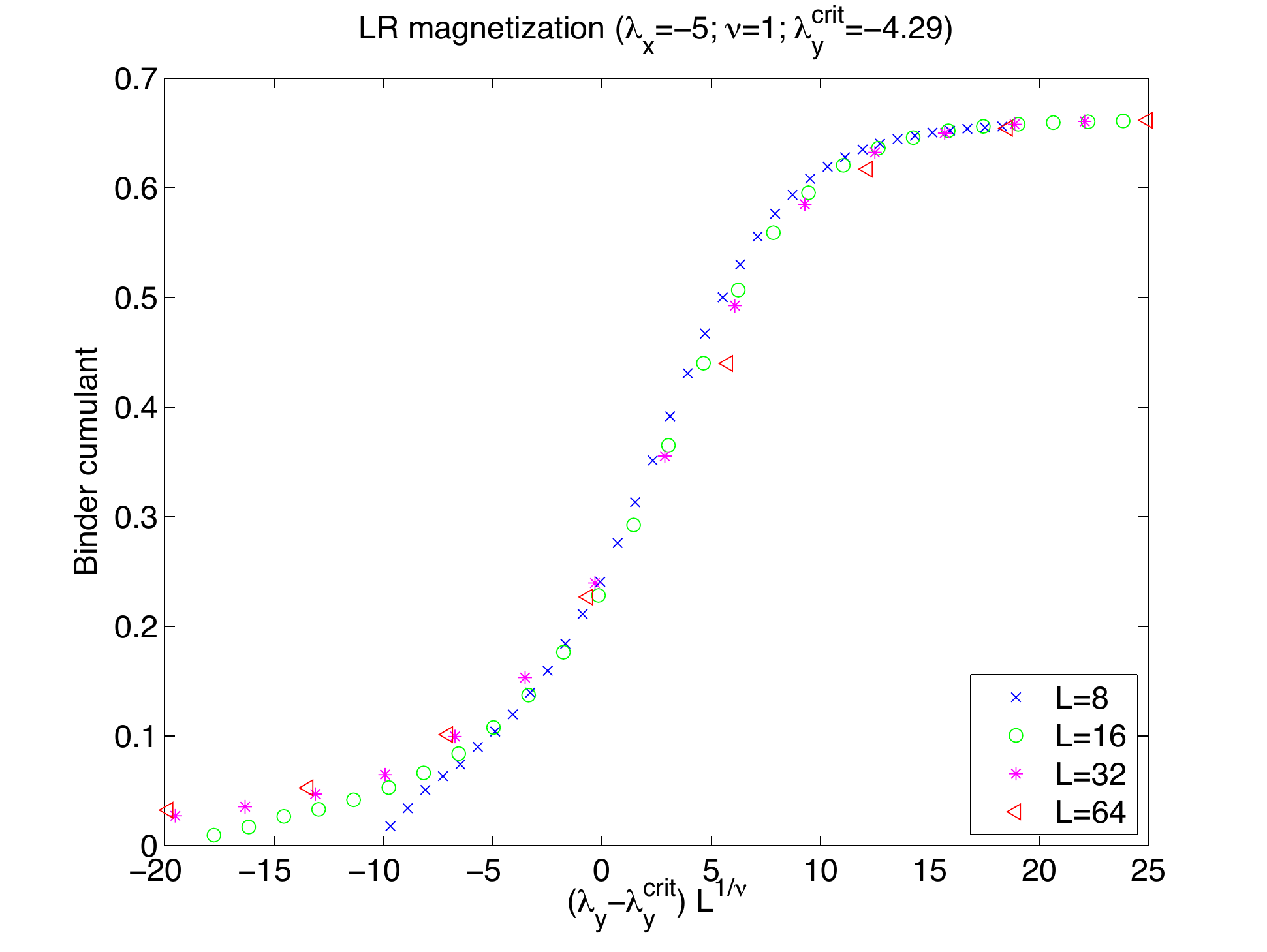}\
\includegraphics[width=0.8\textwidth]{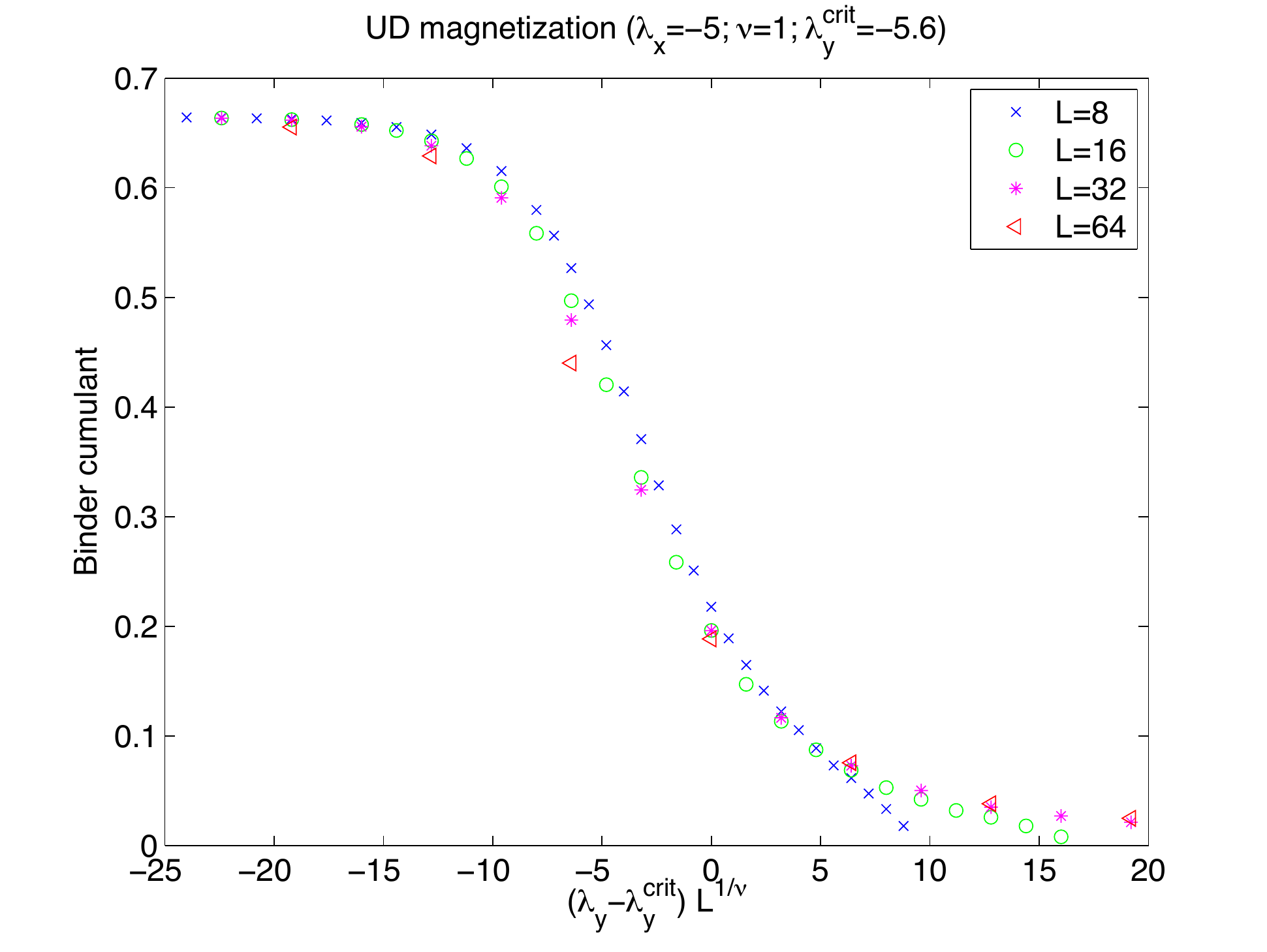}
\end{center}
\caption{\label{fig:exp-nu}Data collapse for the critical exponent $\nu=1$
of the classical two-dimensional Ising model. The $x$ axis is centered around the critical point and rescaled by $L^{1/\nu}$}
\end{figure}

If we rescale the $x$ axis for the Binder cumulant, and plot it as
a function of $\left(\lambda-\lambda_{c}\right)L^{1/\nu}$, then the
data points for all system sizes should collapse. We indeed observe
a good data collapse for the correlation length exponent of the classical
two-dimensional Ising model, $\nu=1$, see Fig.~\ref{fig:exp-nu}.

\end{document}